\documentclass[12pt]{article}
\usepackage{ArxivStyle}

\title{Differentially private sliced inverse regression in the federated paradigm}
\author{Shuaida He\thanks{Department of Statistics and Data Science, Sourthern University of Science and Technology, Shenzhen, China.}, 
	Jiarui Zhang \thanks{Co-first author. Department of Statistics and Data Science, Sourthern University of Science and Technology, Shenzhen, China.}, 
	Xin Chen \thanks{Corresponding author. Department of Statistics and Data Science, Southern University of Science and
		Technology, Shenzhen 518055, China; Email: chenx8@sustech.edu.cn}}
\date{\today}

\begin{document}

\maketitle

\begin{abstract}
	Sliced inverse regression (SIR), which includes linear discriminant analysis (LDA) as a special case, is a popular and powerful dimension reduction tool. In this article, we extend SIR to address the challenges of decentralized data, prioritizing privacy  and communication efficiency. Our approach, named as federated sliced inverse regression (FSIR), facilitates collaborative estimation of the sufficient dimension reduction subspace among multiple clients, solely sharing local estimates to protect sensitive datasets from exposure.  
	To guard against potential adversary attacks, FSIR further employs diverse perturbation strategies, including a novel vectorized Gaussian mechanism that guarantees differential privacy at a low cost of statistical accuracy. 
	Additionally, FSIR naturally incorporates a collaborative variable screening step, enabling effective handling of high-dimensional client data. 
	Theoretical properties of FSIR are established for both low-dimensional and high-dimensional settings, supported by extensive numerical experiments and real data analysis.
\end{abstract}

\section{Introduction}

In recent years,  preserving data ownership and privacy in statistical data analysis has gained significant importance. 
In numerous applications, records carrying sensitive personal information are sourced from diverse origins, posing challenges in aggregating them into a unified dataset for joint analysis. 
This obstacle arises from commonly known transmission restrictions, such as expensive communication costs between a central server and local sites, but it is the privacy concerns that greatly intensify it.
A typical example highlighting this challenge is the collection of patient-level observations from different clinical sites \citep{duan2022heterogeneity, tong2022distributed, liu2022multisite}.

To this end, efficiently extracting valid information from decentralized data while preserving data privacy is critical. Federated learning \citep{mcmahan2017communication, li2020federated}, a widely adopted distributed computing framework, has emerged as a promising approach that emphasizes privacy protection by aggregating partial estimates from different clients to yield a centralized model. 
However, while this computing paradigm preserves client data ownership, it does not directly guarantee record-level privacy, a crucial concern in scenarios lacking a trusted central server or a secure communication environment.
In such cases, adversaries can access statistics calculated by local clients during the uploading procedure \citep{homer2008resolving, calandrino2011you}, 
posing risks of various privacy attacks,  including re-identification, reconstruction, and tracing attacks \citep{bun2014fingerprinting, dwork2015robust, kamath2019privately}, among others.

The tracing attack, also known as membership inference attack, is particularly insidious among those adversaries, as it attempts to identify whether a target individual is a member of a given dataset. While this may appear as a subtle privacy breach, extensive research has demonstrated that even the presence of a single record in a specific dataset can be highly sensitive information \citep{dwork2017exposed}. 
This type of information is closely connected to the concept of differential privacy \citep{dwork2006calibrating}, a rigorous definition of privacy widely adopted in both academia \citep{dwork2009differential, wasserman2010statistical, avella2021privacy} and real-world applications \citep{erlingsson2014rappor, apple2017learning,ding2017collecting, drechsler2023differential}. 
To be specific, let $\mathcal{X}$ be the sample space and $\mathcal{D} \in \mathcal{X}^n$ be a dataset of $n$ records, a randomized algorithm or mechanism $\mathcal{M}: \mathcal{X}^n \rightarrow \mathcal{T}$ guarantees $(\varepsilon, \delta)$-differential privacy if 
\begin{equation}\label{formular:dp}
	P\left(\mathcal{M}(\mathcal{D}) \in \mathcal{S}\right) \leq e^{\varepsilon} P\left(\mathcal{M}\left(\mathcal{D}^{\prime}\right) \in \mathcal{S}\right)+\delta
\end{equation}
for all measurable outputs $\mathcal{S} \subseteq \mathcal{T}$ and all datasets $\mathcal{D}, \mathcal{D}^{\prime}$ that differ in a single record. The parameter set $(\delta, \epsilon)$ controls the level of privacy. 
Intuitively, $\mathcal{M}(\mathcal{D})$ captures the global characteristics of  $\mathcal{D}$, and the goal of privacy is to simultaneously protect every record in $\mathcal{D}$ while releasing $\mathcal{M}(\mathcal{D})$.
A common strategy to construct a differentially private estimator is perturbing its non-private counterpart by random noises \citep{dwork2014algorithmic}. 
Along this line, a variety of fundamental data analyses, such as mean estimation \citep{kamath2019privately}, covariance estimation \citep{biswas2020coinpress}, linear regression \citep{talwar2015nearly}, have been revisited in the context of ($\epsilon,\delta$)-differential privacy protection.
In particular, \citet{cai2021cost} proposed a sharp tracing attack to establish minimax lower bounds for differentially private mean estimation and linear regression.

This paper proposes a federated computation framework for estimating the subspace pursued by a class of supervised dimension reduction methods, termed sufficient dimension reduction (SDR, \citep{cook1994interpretation}), 
with the guarantee of differential privacy. 
Briefly speaking, SDR seeks to replace the original predictors $X=(X_1,\dots,X_p)^{\T}$ in a regression problem with a minimal set of their linear combinations without loss of information for predicting the response $Y$. Mathematically, this can be expressed as 
\begin{equation}\label{formular:sdr}
	Y\indep X|\beta^{\T}X,
\end{equation}
where $\beta \in \mathbb{R}^{p \times d}\ (d<p)$ and $\indep$ indicates independence. 
The subspace spanned by $\beta$, denoted by $\mathcal{S}_{Y|X}\triangleq \operatorname{span}(\beta)$, is the parameter of interest and known as the sufficient dimension reduction subspace, or SDR subspace \citep{cook1996graphics, cook2009regression}. 
Here $d = \dim(\mathcal{S}_{Y|X})$ is often called the structure dimension.
To illustrate the potential risk of privacy breaches in estimating $\mathcal{S}_{Y|X}$, a motivating example is provided.

\begin{example}[A tracing attack on linear discriminant analysis]
	We visit the Body Fat Prediction Dataset \citep{penrose1985generalized}, which includes a response variable BodyFat and $13$ covariates, such as Age, Weight, and Neck circumference, etc. 
	By constructing a binary response $Y=\idop(\operatorname{BodyFat} > 18)$ to indicate whether a man is overweight or not, we can perform linear discriminant analysis (LDA).  Notice that $pr(Y=1) \approx 0.5$ for this dataset.
	To simplify the problem, we transform the covariate $X$ to meet the assumption $X \sim N(0, I_p)$, allowing us to obtain the discriminant direction by \[\beta=E[X \{\idop(Y=1) - \idop(Y=0) \}].\] 
	Given the sensitive information contained in each record, it is desirable to estimate $\beta$ in a differentially private manner.  
	Unfortunately, directly releasing the sample estimate of $\beta$ would compromise individual privacy. 
	To demonstrate this, we propose a tracing attack: 
	$$\mathcal{A}_{\beta}(z_0, \widehat{\beta}(\mathcal{D})) = \langle x_0\{\idop(y_0=1) - \idop(y_0=0) \}  - \beta,  \widehat{\beta}(\mathcal{D})  \rangle,$$
	where $\mathcal{D}=\{z_i: z_i\triangleq (x_i,y_i)\}_{i=1}^{n}$ denotes the dataset for computing $\widehat{\beta}(\mathcal{D})$, $\widehat{\beta}(\mathcal{D})\triangleq \sum_{i=1}^{n}x_i \{\idop(y_{i}=1)-\idop(y_{i}=0)\}/n$, and $z_0 \triangleq (x_0, y_0)$ is a single record to be traced (i.e., to be identified $z_0 \in \mathcal{D}$ or not). 
	
	Clearly, $\mathcal{A}_{\beta}(z_0, \widehat{\beta}(\mathcal{D})) $ takes a large value if $z_0\in \mathcal{D}$ and tends towards zero if $z_0 \notin \mathcal{D}$. This intuitive observation inspires us to treat $\mathcal{A}_{\beta}(\cdot,\widehat{\beta}(\mathcal{D}))$ as a binary classifier that outputs ``in'' if $\mathcal{A}_{\beta}(z_0, \widehat{\beta}(\mathcal{D})) > \tau$ and ``out'' otherwise, given a proper threshold $\tau$. 
	To show the effectiveness of this classifier, we conduct the following experiment. 
	First, since $\beta$ is an unknown parameter, we replace it with $x^{\prime}\{\idop(y^{\prime}=1)-\idop(y^{\prime}=0)\}$, where $(x^{\prime},y^{\prime})$ is a random sample drawn from $\mathcal{D}$. 
	We then randomly divide the remaining samples in $\mathcal{D}$ into two disjoint parts of equal size, $\mathcal{D}_1$ and $\mathcal{D}_2$. As only $\mathcal{D}_1$ will be utilized to calculate $\widehat{\beta}$, we tag the samples in $\mathcal{D}_1$ with the label ``in'' and those in $\mathcal{D}_2$ with the label ``out''. 
	We can then predict the label of each datum $z_0 \in \mathcal{D}_1 \cup \mathcal{D}_2$ by computing $\mathcal{A}_{\beta}(z_0,\widehat{\beta}(\mathcal{D}_1))$. 
	To evaluate the results, we plot the Receiver Operating Characteristic (ROC) curve, which allows us to avoid specifying a particular threshold $\tau$. 
	The red solid line in Figure \ref{fig:demo1} shows the result obtained using the raw $\widehat{\beta}(\mathcal{D}_1)$, which wraps a relatively large area, indicating that $\mathcal{A}_{\beta}$ is an effective classifier, thus is sharp in attacking $\widehat{\beta}(\mathcal{D}_1)$. 
	In contrast, the blue dashed line, representing result obtained using a differentially private $\widetilde{\beta}(\mathcal{D}_1)$ (see Section \ref{sec:method} for details), suggests that $\mathcal{A}_{\beta}$ fails to accurately identify the label of $z_0$ based on $\widetilde{\beta}(\mathcal{D}_1)$, which means that $\widetilde{\beta}(\mathcal{D}_1)$ preserves the privacy of $\mathcal{D}_1$.
	\begin{figure}[!h]
		\centering
		\includegraphics[width=0.5\textwidth]{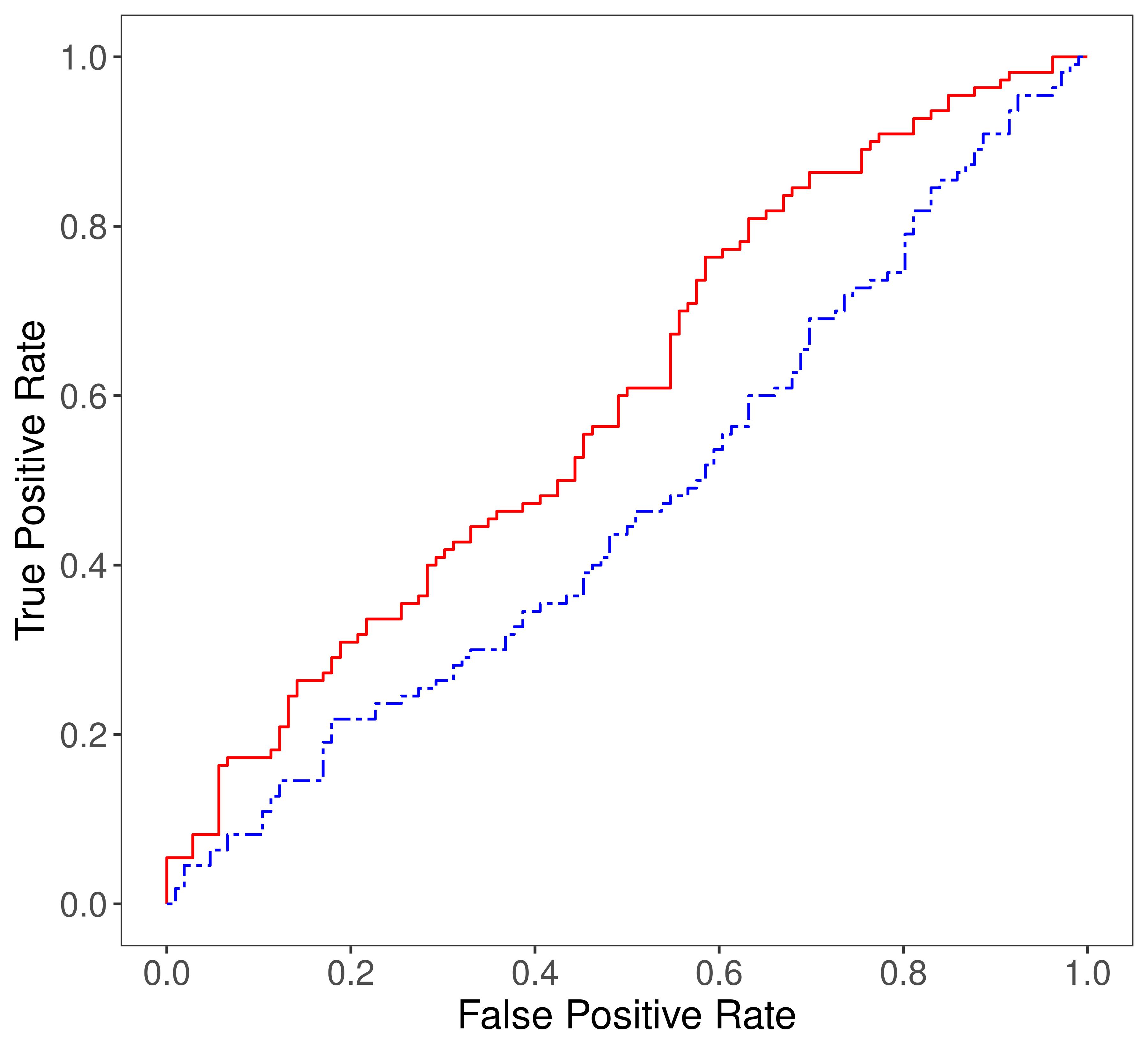}
		\caption{Tracing attack against the discriminant direction of LDA. The ROC curves depict the tracing results based on a naive estimate (red solid line) and a differentially private counterpart (blue dashed line).}
		\label{fig:demo1}
	\end{figure}
\end{example}

The preceding example further reveals the risk of privacy leakage in releasing an unprotected estimate of the basis that spans the SDR  subspace $\mathcal{S}_{Y|X}$. The risk arises because the discriminant direction obtained from LDA is identical to the basis provided by the celebrated sliced inverse regression (SIR, \citep{li1991sliced}), except for scaling \citep{chen2001generalization}. 
In fact, LDA is a special case of SIR when the response is categorical. 
As SIR has played a pivotal role in sufficient dimension reduction, we are motivated to extend it for handling decentralized datasets containing sensitive information. 
Our approach, which we call Federated SIR (FSIR), has the following major contributions.

First, as the name implies, FSIR leverages a federated paradigm to integrate information from diverse clients to estimate the sufficient dimension reduction subspace  $\mathcal{S}_{Y|X}$.  
In contrast to existing distributed SDR methods that are mainly guided by the divide-and-conquer strategy to tackle massive data without considering privacy leakage during communication \citep{xu2022distributed, chen2022distributed, cui2023federated}, our approach prioritizes both client-level (data ownership) and record-level (differential) privacy. 
While similar privacy considerations have been explored in the context of principal component analysis (PCA) \citep{chaudhuri2013near, jiang2016wishart, grammenos2020federated, duchi2022subspace}, an unsupervised counterpart of sliced inverse regression, we are not aware of any literature exploring its extension to sufficient dimension reduction. 
Our work shows that by combining appropriate perturbation strategies, FSIR can guarantee $(\epsilon, \delta)$-differential privacy at a low cost of statistical accuracy. 

Second, we introduce a novel vectorized Gaussian mechanism designed to preserve the information necessary for accurately identifying $\mathcal{S}_{Y|X}$ during the perturbation process. 
This mechanism allows flexible noise direction at the cost of a slightly higher variance. Using a specific algorithm, it generates multivariate noises with a similar eigenspace structure as the signal. Experimental results demonstrate the superiority of this new mechanism over the commonly used i.i.d. Gaussian mechanism.

Third, FSIR can handle high-dimensional data efficiently on each client. 
A large body of literature has contributed to estimating $\mathcal{S}_{Y|X}$ on a unified high-dimensional dataset, including early work on penalized regression-based formulations of SDR methods by \citet{li2007sparse}, the widely-adopted coordinate-independent sparse estimation method (CISE) by \citet{chen2010coordinate}, and the general sparse SDR framework SEAS by \citet{zeng2022subspace}, among others.  
In particular, extensive studies on high-dimensional sliced inverse regression have made notable progress in both methodological development and theoretical understanding \citep{lin2019sparse, tan2018convex, lin2021optimality, tan2020sparse}. 
Despite these advances, an effective strategy for producing a sparse SIR on a decentralized dataset is still lacking. 
Specifically, when dealing with enormous client devices, many of which have limited computation resources, it is desirable to achieve sparsity with a minimal cost on the client-side while leveraging the benefits of federated computing. 
To achieve these goals, we propose a simple yet effective collaborative feature screening method that seamlessly integrates into the FSIR framework to yield a sparse estimator of the basis. Theoretical guarantees for this collaborative screening strategy will be discussed.

Moreover, FSIR is designed to be communication eﬀicient and computational effective.
Compared with multi-round optimization-based approaches \citep{cui2023federated}, FSIR achieves communication efficiency through a one-shot aggregation of local estimates from collaborative clients, thus circumvents introducing noises in each optimization step 
 \citep{abadi2016deep}. 
Moreover, the primary computation cost on each client is attributed to performing singular value decomposition on a small size matrix, leading to computational effectiveness. These characteristics make FSIR well-suited for applications on edge devices, such as smartphones, where efficient and privacy-preserving computations are crucial. 


The following notation will be used in this paper. 
For a positive integer $H$, we write $[H]$ as shorthand for $\{1,\dots,H\}$. 
For a vector $\beta \in \mathbb{R}^p$ and a subset $\mathcal{I} \subseteq [p]$, we use $\beta^{\mathcal{I}}$ to denote the restriction of vector $\beta$ to the index set $\mathcal{I}$. 
Similarly, for $\mathcal{I},\mathcal{J} \subseteq [p]$, $A^{\mathcal{I},\mathcal{J}}$ denotes the $|\mathcal{I}| \times |\mathcal{J}|$ sub-matrix formed by restricting the rows of $A$ to $\mathcal{I}$ and columns to $\mathcal{J}$. 
In addition, for sub-matrix $B = A^{\mathcal{I},\mathcal{J}}$, let $e(B)$ be the embedded matrix into $\mathbb{R}^{p \times p}$ by putting $0$ on entries outside $\mathcal{I} \times \mathcal{J}$.
We write $\operatorname{supp}(\beta):=\left\{j \in[p]: \beta_j \neq 0\right\}$. 
For two vectors $\beta_1$ and $\beta_2$ of same dimension, denote the angle between them by $\angle(\beta_1,\beta_2)$. 
For $x \in \mathbb{R}$ and $R>0$, let $\Pi_{R}(x)$ denote the projection of $x$ onto the closed interval $[-R,R]$.
Define a multi-set as an ordered pair $(\mathcal{B},m)$ where $\mathcal{B}$ is the underlying set formed of distinct elements and $m:\mathcal{B} \rightarrow \mathbb{Z}^{+}$ is the function giving the multiplicity of an element in $\mathcal{B}$. 
For two sets $\mathcal{B}_1$ and $\mathcal{B}_1$, denote their multi-set sum by $\mathcal{B}_1 \biguplus \mathcal{B}_2$. 
For a parameter $\theta \in \Theta$, $\widehat{\theta}$ denotes a raw  estimator and $\widetilde{\theta}$ denotes a differentially private estimator.

\section{Problem Formulation}\label{sec:problem}
\subsection{Preliminaries}\label{sec:problem-pre}
We revisit the concept of differential privacy before moving on. Intuitively, the privacy level set $(\delta, \epsilon)$ in definition (\ref{formular:dp}) measures how much information $\mathcal{M}(\cdot)$ reveals about any individual record in the dataset $\mathcal{D}$, where the smaller value of $\epsilon$ or $\delta$ imposes the more stringent privacy constraint.  
This definition leads to some nice properties of differential privacy. The two most useful ones are listed as follows, which provide a convenient way to construct complex differentially private algorithms.
\begin{property}[Post-processing property, \citet{dwork2006calibrating}]
	\label{prop:DP-pp}
	If $\mathcal{M}: \mathcal{X}^n \rightarrow \Theta_0$ is $(\epsilon,\delta)$-differentially private and $\mathcal{M}^{\prime}: \Theta_0 \rightarrow \Theta_1$ is any randomized algorithm, then $\mathcal{M}^{\prime} \circ \mathcal{M}$ is $(\epsilon,\delta)$-differentially private.
\end{property}
\begin{property}[Composition property, \citet{dwork2006calibrating}]
	Let $\mathcal{M}_i$ be $(\epsilon_i,\delta_i)$-differentially private for $i = 1, 2$, then $\mathcal{M}_1 \circ \mathcal{M}_2$ is $(\epsilon_1 + \epsilon_2,\delta_1+ \delta_2)$-differentially private.
\end{property} 

A standard approach for developing differentially private algorithms involves introducing random noises to the output of their non-private counterparts, see \citep{dwork2014algorithmic}. One such technique, known as the Gaussian mechanism, employs independently and identically distributed (i.i.d.) Gaussian noise. The noise's scale is governed by the $l_2$ sensitivity of the algorithm, which quantifies the maximum change in the algorithm's output resulting from substituting a single record in its input data set. The formal definition is presented below.

\begin{definition}[$l_{2}$-sensitivity]\label{def:sen}
	For an algorithm $f:\mathcal{D} \rightarrow \mathbb{R}^p$, its $l_2$-sensitivity is defined as 
	\[\Delta_{2}(f)=\sup_{\mathcal{D},\mathcal{D}^{\prime}} ||f(\mathcal{D}) - f(\mathcal{D}^{\prime})||_2,\]
	where $\mathcal{D}, \mathcal{D}^{\prime} \in \mathcal{X}^n$ are datasets differing in a single record. 
\end{definition}

\begin{definition}[Gaussian mechanism]
	\label{def:gm}
	For any vector-valued algorithm $f: \mathcal{D} \rightarrow \mathbb{R}^p$ which satisfies $\Delta_2(f) < \infty$, the Gaussian mechanism is defined as $$\widetilde{f}(\mathcal{D}) \triangleq f(\mathcal{D})+\xi,$$
	where $\xi \triangleq (\xi_1,\xi_2,\dots,\xi_p)$ and $\xi_j, \ j \in [p]$ is an i.i.d. sample drawn from $\mathcal{N}(0, \sigma_{\xi}^2)$.
\end{definition}

Next we briefly review sliced inverse regression. 
Let $Y \in \mathbb{R}$ be a response variable and $X=(X_1,\dots,X_p)^\T \in \mathbb{R}^p$ be the associated covariates, with  covariance matrix $\Sigma=\cov(X)$. 
Assume the SDR subspace $\mathcal{S}_{Y|X} = \operatorname{span}(\beta)$, where $\beta\in\mathbb{R}^{p\times d}$ is the basis matrix and $d\ (d < p)$ is the structure dimension.  
SIR seeks for an unbiased estimate of $\mathcal{S}_{Y|X}$. 
Without loss of generality, assume $Y$ is categorical with $H$ fixed classes, denoted as $Y \in [H]$; when $Y$ is numerical, we simply follow \citet{li1991sliced} to construct a discrete version $\widetilde{Y}$ of $Y$ by partitioning its range into $H$ slices without overlapping. 
This discretization operation will not lose information for estimating SDR subspace, i.e., $\mathcal{S}_{\widetilde{Y}|X} = \mathcal{S}_{Y|X}$, given $H$ sufficiently large \citep{bura2001extending}, and we refer to a value $h \in [H]$ as a slice or a class.
In preparation, let $p_h \triangleq \mathrm{pr}(Y = h)$ and $m^0_h \triangleq E\{X-E(X) |Y = h\}$, the kernel matrix is then given by 
\[\Lambda^0 \triangleq \sum_{h=1}^{H}p_h m^{0}_{h} m^{0\T}_{h}.\]
\citet{li1991sliced} has shown that, if $X$ satisfies the so-called linear conditional mean (LCM) condition, i.e., $E(X \mid \beta^\T X)$ is a linear function of $\beta^\T X$, then $\Sigma^{-1} m^0_h \in \mathcal{S}_{Y \mid X}$ for $h \in [H]$.
This result implies that $\Sigma^{-1}\operatorname{col}(\Lambda^0) \subseteq \mathcal{S}_{Y|X}$. 
In practice, we further assume the coverage condition, that is,  $\Sigma^{-1}\operatorname{col}(\Lambda^0) = \mathcal{S}_{Y|X}$, so the first $d$ eigenvectors of $\Sigma^{-1}\operatorname{col}(\Lambda^0)$ spans $\mathcal{S}_{Y|X}$ and the remaining $p-d$ eigenvalues equal to zero. 

At the sample level, we can compute the plug-in estimator $\widehat{\Sigma}$ and $\operatorname{col}(\widehat{V}_{H}^{0})$ to estimate $\mathcal{S}_{Y|X}$, where $\widehat{V}_{H}^{0}$ is the matrix formed by the top $d$ principal eigenvectors of $\widehat{\Lambda}^{0}$. 
\citet{lin2018consistency} has shown that the space spanned by $\widehat{\Sigma}^{-1}\widehat{V}_{H}^{0}$ yields a consistent estimate of $\mathcal{S}_{Y|X}$ if and only if $\rho_{n}:=\frac{p}{n} \rightarrow 0$ as $n \rightarrow \infty$, when the slice number $H$ is a fixed integer independent of $n$ and $p$, and the structure dimension $d$ is bounded.

\subsection{Computing SIR in a federated paradigm}\label{sec:problem-fed}
Consider a dataset $\mathcal{D} \triangleq \{\mathcal{D}^{(1)}, \cdots, \mathcal{D}^{(K)} \}$ distributed across $K$ clients, 
where $\mathcal{D}^{(k)}=\{(x_i,y_i): x_i \in \mathbb{R}^p, y_i \in [H]\}_{i=1}^{n_k}$ is the subset stored on the $k$th client with $n_k$ samples, $k \in [K]$. Let $N=\sum_{k=1}^{K}n_k$ be the total size of $\mathcal{D}$. 
Generally, we do not require $n_k > p$ for any single client $k \in [K]$, thus on some or even all clients, data may follow a high-dimensional  setting ($p > n_k$, or $p \gg n_k$). Meanwhile, given $K$ sufficiently large, it is also reasonable to assume a low-dimensional setting for the whole dataset $\mathcal{D}$, that is, $p < N$.

To handle this decentralized dataset, we need modify the classical SIR method slightly. 
Denote $m_h \triangleq E[\{X-E(X)\} \idop(Y = h)]$, we immediately have 
$\Sigma^{-1} m_h \in \mathcal{S}_{Y \mid X}$
for $h \in [H]$, since $m_h = p_h m^0_{h}$. 
For convenience of expression, define the slice mean matrix $M \triangleq (m_1,\dots,m_H) \in \mathbb{R}^{p \times H}$ and the kernel matrix $\Lambda \triangleq MM^{\T}$, then
\begin{equation}\label{formular:sir}
	\Sigma^{-1}\operatorname{col}(\Lambda) \subseteq \mathcal{S}_{Y|X}.
\end{equation}

Using $\Lambda$ instead of $\Lambda^0$ is beneficial for our federated computation. 
The first advantage arises in scenarios with imbalanced classes, where a client only has a small portion of observations for certain classes. Estimating $\Lambda^0$ in such cases can be highly unstable, resulting in a biased global estimator. 
This situation is frequently encountered in streaming data, as discussed by \citet{cai2020online} in developing their online SIR estimator. 
In our problem, utilizing $\Lambda$ further allows us to eliminate privacy leakage in uploading $\widehat{p}_h$'s, thus avoiding unnecessary complexities in both algorithm design and theoretical analysis. 
Additionally, as detailed in Section \ref{sec:method}, $\Lambda$ has a smaller $l_2$-sensitivity, reducing the noise scale required for perturbation strategies. 

\begin{algorithm}[!ht]
	\caption{The FSIR framework.}	
	\label{alg:fsir}
	\KwData{$\mathcal{D}=\bigcup_{k=1}^{K}\mathcal{D}^{(k)}$}
	\KwIn{Clients number $K$, privacy level sets $(\epsilon_x, \delta_x)$ and $(\epsilon_m, \delta_m)$}.
	\SetKwProg{LoopRounds}{In round}{:}{end}
	\SetKwProg{ClientsDo}{Client}{ do in parallel}{end}
	\SetKwProg{ServerDo}{Server}{ do}{end}
	
	\ClientsDo{$k \in [K]$}{
		1.\If{$p>n_k$}{
			/*\textit{\textbf{{\footnotesize An alternative step to tackle high-dimensional data.}}}*/\\
			Estimate local active variable set $\mathcal{A}^k \leftarrow \operatorname{CCMD-Filter}$ (See Algorithm  \ref{alg:fed-screening} in Section \ref{sec:sparse})\;
			Upload $\mathcal{A}^{(k)}$ and pull $\mathcal{A}$ from Server\;
			$X^{(k)} \leftarrow X^{(k)}_{\mathcal{A}}$\;
		}
		2. Upload the private slice mean matrix $\widetilde{M}^{(k)}$ (See Algorithm \ref{alg:dp-vmg} in Section \ref{sec:vgm})\; 
		3. Upload the private covariance matrix $\widetilde{\Sigma}^{(k)}$ as well as the sample size $n_{k}$\;
	}
	\ServerDo{}{
		1. \If{$\mathcal{A}^{(k)} \neq \emptyset$}{
			/*\textit{\textbf{\footnotesize An alternative step to tackle high-dimensional data.}}*/\\
			Update $\mathcal{A} \leftarrow \operatorname{CCMD-Filter}$ (See Algorithm \ref{alg:fed-screening} in Section \ref{sec:sparse})\;
		}
		2. Merge $\widetilde{\Sigma}= \sum_{k=1}^{K} n_{k} \widetilde{\Sigma}^{(k)}/\sum_{k=1}^{K} n_{k}$\;
		3. Merge $\widetilde{M} = \sum_{k=1}^{K} n_{k} \widetilde{M}^{(k)}/\sum_{k=1}^{K} n_{k}$ to obtain $(\widetilde{U},\widetilde{S},\widetilde{V}) \leftarrow\operatorname{SVD}(\widetilde{M})$ \;
		4. Return the global estimate $\widetilde{\beta}=\widetilde{\Sigma}^{-1}\widetilde{U}$.\\
	}
\end{algorithm}

Algorithm \ref{alg:fsir} presents a federated framework for computing SIR on the decentralized dataset $\mathcal{D}$, catching a glimpse of our FSIR method. In this framework, each client $k \in [K]$ operates independently to estimate the slice mean matrix as $\widehat{M}^{(k)}$ and the covariance matrix as $\widehat{\Sigma}^{(k)}$ based on their local dataset $\mathcal{D}^{(k)}$, without needing to access external data. 
Subsequently, these two matrices are uploaded to the central server after a perturbation operation to ensure differential privacy  (see details later). 
A global estimate of the SDR subspace is obtained on the server in Step 4, achieved through the merging of $\widetilde{M}^{(k)}$ and $\widetilde{\Sigma}^{(k)}$ for all $k \in [K]$, where $\widetilde{M}^{(k)}$ and $\widetilde{\Sigma}^{(k)}$ denoting the perturbed matrices.

At first glance, the pooling and averaging operations at Step $2$ and $3$ on the server might suggest that FSIR is merely another application of the FedAvg algorithm \citep{mcmahan2017communication}. 
However, FSIR distinguishes itself by emphasizing on one-shot communication and, more importantly, privacy protection. 
Throughout the procedure, each client only computes and uploads local model estimates to the server, ensuring that  sensitive records would remain localized with their owner. In this way, FSIR restricts the central server or any other client $j$ can only probe the dataset of client $k$ through its uploaded estimates. 
Meanwhile, admitting the existence of potential malicious participants who may conduct tracing attacks on the shared estimates, FSIR necessitates the differential privacy of both $\widetilde{M}^{(k)}$ and $\widetilde{\Sigma}^{(k)}$. 
This privacy requirement ensures that any operation performed on these estimates does not cause additional privacy breaches, thanks to the post-processing property of differential privacy.

An essential step in Algorithm \ref{alg:fsir} involves the construction of differentially private $\widetilde{\Sigma}^{(k)}$ and $\widetilde{M}^{(k)}$ for each client $k$. 
Privacy-preserving estimation of covariance matrices has been extensively studied, as evidenced by a large body of literature \citep{chaudhuri2012near,grammenos2020federated, biswas2020coinpress, wang2021differentially}. 
While it is not the primary focus of this paper, we simply adapt the approach suggested by \citet{chaudhuri2012near} within our framework, corresponding to Step $3$ on the client. 
The privacy of corresponding operation is guaranteed by Lemma \ref{lem:dp-cov}.  
In the next section, our attention shifts towards the development of a differentially private slice mean matrix, whose singular subspace plays a crucial role in estimating $\mathcal{S}_{Y|X}$.

\begin{lemma}\label{lem:dp-cov}\citep{chaudhuri2012near}
	Let $X \in \mathbb{R}^{p \times n}$ be the centralized design matrix on the client with $\|X_i\|\leq 1$ for all $i \in [n]$. Denote  $\widehat{\Sigma}=\frac{1}{n} X X^{\T}$.
	Let $A \in \mathbb{R}^{p \times p}$ be a symmetric random matrix whose entries are i.i.d. drawn from $N(0,\sigma_x^2)$, where 
	\[ \sigma_x=\frac{p+1}{n \epsilon_x} \left\{2 \log \left(\frac{p^2+p}{2 \sqrt{2 \pi}\delta_x}\right)\right\}^{1/2}+\frac{1}{n {\epsilon_x}^{1/2}}. \]
	Then $\widetilde{\Sigma} \triangleq \widehat{\Sigma} + A$  is $(\epsilon_x,\delta_x)$-differentially private.
\end{lemma}

Notice the condition $\|X_i\| \leq 1$ is not commonly encountered in the realm of sufficient dimension reduction. 
We address this by devising a strategy to perturb $\widehat{\Sigma}/c_R^2 = \frac{1}{n}(X/c_R)(X/c_R)^{\top}$  in accordance with the magnitudes suggested in Lemma \ref{lem:dp-cov}. Here, $c_R$ denotes the maximum norm $\max_i\|X_i\|$, which is controlled by the truncation level $R$ (see Section \ref{sec:iid}). 
Notably, this scaling transformation has no impact on the estimation of $\mathcal{S}_{Y|X}$.


\section{Estimation with differential privacy}\label{sec:method}
\subsection{Private slice mean matrix: a preliminary way}\label{sec:iid}
On the client-side, a natural approach for constructing a differential private slice mean matrix  is to apply the i.i.d. Gaussian mechanism to the mean vector of each slice; see Proposition \ref{th:iid-dp}, which can be easily derived from Theorem A.1 in \citet{dwork2014algorithmic}. 
\begin{proposition}\label{th:iid-dp}
	Let $\{(X_i^\T,y_i)^{\T} \in \mathbb{R}^{p+1}, i \in [n]\}$ be i.i.d. samples on a client.  
	Denote $\widehat{m}_{h} \triangleq \frac{1}{n}\sum_{i=1}^{n} X_i \idop(y_i=h)$ and assume its $l_2$-sensitivity $\Delta_{2} < \infty$. 
	Define $\widetilde{m}_h \triangleq \widehat{m}_h + (\xi_1,\xi_2,\dots,\xi_p)^{\T} \in \mathbb{R}^{p}$, 
	where $\xi_j \sim N(0,\frac{ 2 \Delta_{2}^2 \log(1.25/\delta)}{\epsilon^2})$, then $\widetilde{M} \triangleq (\widetilde{m}_1,\dots,\widetilde{m}_H) \in \mathbb{R}^{p \times H}$ is $(\epsilon,\delta)$-differential private.
\end{proposition}

Clearly, the noise scale is determined by both of the privacy level set $(\epsilon,\delta)$ and the $l_2$-sensitivity $\Delta_{2}$. 
Recall that sliced inverse regression usually assumes that $X$ is integrable and has an elliptical distribution to satisfy the LCM condition, which does not guarantee $\Delta_{2} < \infty$ in general. Therefore, we truncate $X$ to restrict it on a finite support.
Given the truncation level $R$, let \[\overline{m}_{h,j}=\frac{1}{n} \sum_{i=1}^{n}\Pi_{R}(X_{ij})\idop(y_i=h),  j \in [p].\]
Denote $\overline {m}_{h}=(\overline{m}_{h,1},\dots,\overline{m}_{h,p})^{\T}$ and $\overline{M} \triangleq (\overline {m}_{1},\dots,\overline {m}_{H}) \in \mathbb{R}^{p \times H}.$
We then utilize $\overline {M}$ rather than $\widehat{M}$ to meet the assumption $\Delta_{2} < \infty$. 
Since a sample can be only located in one slice, we easily obtain $\Delta_{2}=2R\sqrt{p}/n$ over a pair of data sets which only differ by one single entry, where $n$ is the client sample size.

The differential privacy is always guaranteed at the expense of estimation accuracy \citep{wasserman2010statistical, bassily2014private}.  In real applications, a client could have extremely limited observations, causing the scale of added noises  significantly larger than of the signal. For this, we suggest to set an upper bound $\sigma_0^2$ as a tolerated scale of noise and calculate the minimal sample size  
\[n_{\epsilon,\delta,p,R}=\frac{2R\sqrt{2p\log(1.25/\delta)}}{\sigma_{0}\epsilon}\] 
required by each client. Thus for client $k  \in [K]$, only if $n_k \ge n_{\epsilon,\delta,p,R}$, we compute and upload $\widetilde{M}_k$ following Proposition \ref{th:iid-dp}.
 
 \begin{remark}
 	The truncating operation is widely adopted for both theoretical studies and practical applications, we refer to \citep{lei2011differentially, cai2021cost} for more details on choosing a proper truncation level. As for the selection of privacy level set, we adopt the convention $\epsilon = O(1)$ and $\delta = o(1/n)$, which is usually considered as the most permissive setting to provide a nontrivial privacy protection \citep{dwork2014algorithmic}.
 \end{remark}


\subsection{Private slice mean matrix: singular space oriented}\label{sec:vgm} 
The use of vanilla Gaussian mechanism in the previous section is intuitive and simple, yet has a significant limitation: it does not account for the impact of perturbation operation on the left singular space of $\overline{M}$, which is crucial for estimating $\mathcal{S}_{Y|X}$, as revealed by (\ref{formular:sir}). To mitigate this, we propose a novel perturbation strategy, termed vectorized Gaussian (VGM) mechanism, to provide $(\epsilon,\delta)$-differential privacy for $\overline{M}$ without sacrificing too much information we needed.
We start by giving its definition.

\begin{definition}[Vectorized Gaussian mechanism]
	\label{def:vgm}
	For any vector-valued algorithm $f: \mathcal{D} \rightarrow \mathbb{R}^p$ which satisfies $\Delta_2(f) < \infty$, define the vectorized Gaussian mechanism as \[\widetilde{f}(\mathcal{D}) \triangleq f(\mathcal{D})+\xi,\]
	where $\xi \sim \mathcal{N}_{p}(0,\Sigma_{\xi})$. 
\end{definition}

Obviously, the noise vector $\xi$ is characterized by the covariance matrix $\Sigma_{\xi}$. In particular, if $\Sigma_{\xi}=\{2\Delta_{2}^2 \log(1.25/\delta)/\epsilon^2\} I_{p}$, the VGM mechanism coincides with the i.i.d. Gaussian mechanism. 
Theorem \ref{th:dp-vgm} provides a sufficient condition for holding the $(\epsilon,\delta)$-differential privacy of VGM mechanism.

\begin{theorem}\label{th:dp-vgm}
	Let $\{(X_i^\T,y_i)^{\T} \in \mathbb{R}^{p+1}, i \in [n]\}$ be i.i.d. samples on a client.  
	Denote $\widehat{m}_{h} \triangleq \frac{1}{n}\sum_{i=1}^{n} X_i \idop(y_i=h)$ and assume its $l_2$-sensitivity  $\Delta_{2} < \infty$. 
	Define $\widetilde{m}_h \triangleq \widehat{m}_h + \xi \in \mathbb{R}^{p}$, 
	where $\xi \sim \mathcal{N}_p(0, \Sigma_{\xi})$. Suppose the eigenvalue decomposition $\Sigma_{\xi}=U_{\xi} V_{\xi}U_{\xi}^{\T}$. Then $\widetilde{M} \triangleq (\widetilde{m}_1,\dots,\widetilde{m}_H) \in \mathbb{R}^{p \times H}$ is $(\epsilon,\delta)$-differentially private, if $\Sigma_{\xi}$ satisfies 
	\[
	\|\Sigma_{\xi}^{-1}\|_2 \leq \frac{4\log{\frac{2}{\delta}} + 2\epsilon - 4\{(\log{\frac{2}{\delta}})^2 + \log{\frac{2}{\delta}}\epsilon\}^{1/2}}{\Delta_{2}^2}.
	\]
\end{theorem}

By Taylor expansion and some algebraic computations, we can further obtain
\[
\frac{4\log{\frac{2}{\delta}} + 2\epsilon - 4\{(\log{\frac{2}{\delta}})^2 + \log{\frac{2}{\delta}}\epsilon\}^{1/2}}{\Delta_2^2} \approx \frac{\epsilon^2}{2\Delta_2^2\log{\frac{2}{\delta}}}.
\]
For expression convenience, we assume the client sample size $n \geq  n_{\epsilon,\delta,p,R}$ and use 
\begin{equation}\label{formula:upperbound} 
	\sigma^2_{vgm} \triangleq \frac{n^2 \epsilon^2}{8 R^2 p \log{\frac{2}{\delta}}}
\end{equation}
as an approximate upper bound of $\|\Sigma_{\xi}^{-1}\|_2$.  
We further denote the diagonal elements of $V_{\xi}$ by $v_1,\dots,v_p$, which are arranged in decreasing order.

Theorem \ref{th:dp-vgm} presents an inspiring result that allows for flexible design of the eigenvectors $U_{\xi}$ while achieving differential privacy by bounding $s_p$.  
Building upon this finding, we proceed by performing singular value decomposition on $\overline{M}$, yielding $\overline{M}=W_1 S W_2^{\T}$. 
We take $U_{\xi} = W_1$ and sort the diagonal elements of $V_{\xi}$ in the same order as the diagonal elements of $S$, while enforcing the constraint $v_p \geq \sigma^2_{vgm}$.
The key idea here is to construct a $p$-dimensional noise vector $\xi$ with a specific covariance structure, such that its eigenspace aligns with the left singular subspace of the slice mean matrix $\overline{M}$. 
Thus the perturbed matrix $\widetilde{M}=\overline{M}+E_0$ can preserve the relevant information about $\mathcal{S}_{Y|X}$, where $E_0$ is a noise matrix whose columns are generated from $\xi$.

Algorithm \ref{alg:dp-vmg} outlines more details about our strategy. 
Steps $1$ to $3$ split the left singular space of $\overline{M}$ into two parts: a $d$-dimensional subspace spanned by the leading singular vectors and its orthogonal complement. 
Since the first part will be utilized to construct $\mathcal{S}_{Y|X}$, we preserve these eigenvectors and maintain their ordering based on corresponding eigenvalues. 
Meanwhile, we propose that the eigenvectors in the second part have no significant difference thus share the same eigenvalue.
Following this, Steps $4$ construct a matrix $V$ by appropriately arranging diagonal elements, which leads to $\Sigma_{\xi}=W_1 V W_1^{\T}$ and $\|\Sigma_{\xi}^{-1}\|_2 =\sigma^2_{vgm}$.  
Finally, Steps $5$-$6$ generate noise vectors and upload perturbed slice mean matrix.

\begin{algorithm}[!ht]
	\caption{Compute private slice mean matrix via the VGM mechanism.}
	\label{alg:dp-vmg}		
	\KwData{$\mathcal{D}=\{(X_i,y_i) \in \mathbb{R}^{p+1}, i \in [n]\}$.}
	\KwIn{Privacy level set $(\epsilon, \delta)$,  truncation level $R$.}
	\KwOut{Perturbed slice mean matrix $\widetilde{M}\in \mathbb{R}^{p \times H}$.}
	\SetKwProg{LoopLabels}{For label}{:}{end}
	\SetKwFunction{TruncatedCMean}{TruncatedConditionalMean$_{R}$}
	\SetKwProg{Fun}{Function}{:}{end}
	1. Compute $\sigma^2_{vgm}$ and the truncated slice mean matrix $\overline{M}$\;
	2. Obtain $(W_1,S,W_2^{\T}) \leftarrow \operatorname{SVD}(\overline{M})$, where $S \in \mathbb{R}^{p \times p}$. Denote the $j$th diagonal element of $S$ by $s_j,\ j\in[p]$\;
	3. Compute the eigengap $r_j=s_{j+1}-s_j$ for $j\in[p-1]$ and estimate the structure dimension $d$ by ranking $r_j$'s\;
	4. Construct a diagonal matrix $V \in \mathbb{R}^{p\times p}$ with the $j$th diagonal element $v_j$ equals to $\sigma^2_{vgm} + r_j$ for $1\leq j\leq d$ and $\sigma^2_{vgm}$ for $d+1 \leq j \leq p$\;	
	5. Generate a matrix $Z \in \mathbb{R}^{p \times H}$, where $Z_{j,h} \sim_{i.i.d.} N(0,1)$ for $ j \in [p],\ h \in [H]$\;
	6. Obtain $E_0 \triangleq W_1 V^{1/2} Z \in \mathbb{R}^{p\times H}$ and upload $\widetilde{M} \triangleq \overline{M} + E_0$.\\
\end{algorithm}

\begin{remark}
	Step $3$ in Algorithm \ref{alg:dp-vmg} estimates the structure dimension $d$ by simply ranking the eigengaps of $\overline{M}$. 
	While this approach circumvents additional computational burden, alternative methods exist for determining the structure dimension. We refer interested reader to \citep{luo2021order} for a detailed study.
\end{remark}

\begin{remark} 
	Compared with the i.i.d. Gaussian mechanism, the VGM mechanism preserves the information carried by the left singular subspace of $\overline{M}$ at the cost of a higher variance of noise. 	
	Notice we have \[ \|\Sigma_{\xi}\|_2 \geq \frac{8 R^2 p \log{\frac{2}{\delta}}}{n^2 \epsilon^2}\]
	and the minimum	is reached if all the eigenvalues of $\Sigma_{\xi}$ are the same. In practice, we can fix the scale of $\|\Sigma_{\xi}\|_2$ to be ${C R^2 p \log{\frac{2}{\delta}}}/({n^2 \epsilon^2})$, where the constant $C \geq 8$. 
	To prevent the noise from overpowering the signal,  the client sample size should be sufficiently large.  Specifically, we require
	\[ \frac{C R^2 p \log{\frac{2}{\delta}}}{n^2 \epsilon^2} = o(1)\]
	for our theoretical investigation.
\end{remark}

\subsection{Sparse estimation in the high-dimensional setting}\label{sec:sparse}
In many scenarios, client data can be high-dimensional ($p>n$) or even ultra-high-dimensional ($p=o(\exp(n^\alpha))$). 
Since SIR relies on $p/n \rightarrow 0$ as $n \rightarrow \infty$ to guarantee the consistency of $\widehat{V}_{H}$ \citep{lin2018consistency}, we introduce a feature screening procedure as an initial step to handle these data. 
Recall that SIR is developed upon the observation that $\Sigma^{-1} \{ E(X|Y) - E(X)\} \in \mathcal{S}_{Y \mid X}$, implying that  $\mathcal{S}_{Y|X}$ degenerates if $E(X|Y) - E(X)=0$ almost surely. 
This motivates us to define the active set as \[\mathcal{T}=\{ j \in [p] \mid E(X_j|Y) \neq E(X_j) \ \mathrm{a.s.}\}\] 
and utilize the criterion \[\omega_{j,h} \triangleq |E\{X_j - E(X_j) |Y = h\}|, \ h \in [H]\] to identify $\mathcal{T}$.
To take advantage of the decentralized data, we compute $\omega_{j,h}^{(k)}$ on each client $k \in [K]$ and vote for screening on the server. The entire procedure is designed to operate in a federated paradigm and we call it the Collaborative Conditional Mean Difference (CCMD) filter.   We clarify its implementation details in Algorithm \ref{alg:fed-screening}.
With the inclusion of the CCMD filter, our FSIR framework now successfully handles high-dimensional data and is fully operational. 

\begin{algorithm}[!ht]
	\KwData{$\mathcal{D}=\bigcup_{k=1}^{K}\mathcal{D}^{(k)}$.}
	\KwIn{slice number $H$, client number $K$, threshold $t$.}
	\KwOut{Estimated active set $\widehat{\mathcal{T}}$.}
	
	\SetKwProg{LoopLabels}{For label}{:}{end}
	\SetKwProg{ClientsDo}{Client}{ do in parallel:}{end}
	\SetKwProg{ServerDo}{Server}{ do:}{end}
		\ClientsDo{$k \in [K]$}{
			1. Compute $\Omega^{(k)} \in \mathbb{R}^{p\times H}$ where $\Omega_{j,h}^{(k)} \triangleq \widehat{w}_{j,h}^{(k)}$\;
			2. \LoopLabels{$h \in [H]$}{
				Estimate the active set on $h$th slice  $\widehat{\mathcal{T}}^{(k)}_{h}\triangleq \{j \in [p] \mid  \Omega_{j,h}^{(k)} > t\}$\;
			}
			3. Upload the $k$th multiset $\mathcal{B}^{(k)}\triangleq \biguplus_{ h \in [H]} \widehat{\mathcal{T}}^{(k)}_{h}$\;
		}
		\ServerDo{}{
			4. Obtain the global multiset $\mathcal{B}\triangleq \biguplus_{k \in [K]} \mathcal{B}^{(k)}$\;
			5. Return the global active set $\widehat{\mathcal{T}}\triangleq \{j \in \mathcal{B} \mid \operatorname{multiplicity}(j) > \frac{K}{2}\}$.\\
		}
	\caption{The Collaborative Conditional Mean Difference (CCMD) filter.}
	\label{alg:fed-screening}
\end{algorithm}

\begin{remark}
	Step 3 in Algorithm \ref{alg:fed-screening} does not involve privacy leakage, since this operation only uploads an index set of possible active variables rather than their mean values. For the later case, \citet{dwork2018differentially} has developed a differentially private approach, named ``Peeling'', to estimate the top-$s$ largest elements in a $p$-dimensional mean vector.
\end{remark}

\section{Theoretical analysis}\label{sec:theory}
We focus on the homogeneous scenario, where all clients share the same SDR subspace.
The heterogeneous case is left for future research due to its additional technical challenges. 
For convenience, we define the central curve $m(Y) \triangleq {E}(X|Y)$ and assume $E(X) = 0$. 
Suppose sample sizes across all slices in the decentralized dataset $\mathcal{D}$ are equal, denoted by $n_{0} = N / H$.
The following technical conditions are required for our theoretical studies. 

\begin{condition}\label{con1}
For any $\beta \in \mathbb{R}^{p \times d}$, $E(X|\beta^{\T}X)$ is a linear combination of $\beta^{\T}X$.
\end{condition}
\begin{condition}\label{con2}
The rank of $\var\{m(Y)\}$ equals the structure dimension $d$. 
\end{condition}
\begin{condition}\label{con3}
$X$ is sub-Gaussian and there exist positive constants $C_1$, $C_2$ such that 
\begin{align*}
C_1 \leq \lambda_{\mathrm{min}}(\Sigma)\leq \lambda_{\mathrm{max}}(\Sigma)\leq C_2,
\end{align*}
where $\lambda_{\mathrm{min}}(\Sigma)$ and $\lambda_{\mathrm{max}}(\Sigma)$ are the minimal and maximal eigenvalues of $\Sigma$ respectively.
\end{condition}
\begin{condition}\label{con4}
 Assume $m(Y)$ has finite fourth moment and is $\nu$-sliced stable (See definition \ref{stable}).
\end{condition}

\begin{definition}\label{stable}
	For two positive constants $\gamma_1 <1 <\gamma_2$, let $\mathcal{A}_{H}(\gamma_1,\gamma_2)$ be the collection of all the partition $-\infty = a_0 < a_1 < \cdots < a_H = \infty$ of $\mathbb{R}$ satisfying that 
	\begin{align*}
		\frac{\gamma_1}{H} \leq \pr(a_i \leq Y < a_{i+1}) \leq \frac{\gamma_2}{H}.
	\end{align*}
	Then $m(Y)$ is called $\nu$-sliced stable with respect to $Y$ for some $\nu > 0$ if there exist positive constants $\gamma_i$, $i = 1,2,3$ such that for any $\alpha \in \mathbb{R}^p$ and any partition in $\mathcal{A}_{H}(\gamma_1, \gamma_2)$, 
	\begin{align*}
		\frac{1}{H}|\sum_{h=0}^{H-1}\var{\left\{\alpha^{\T}m(Y)|a_h \leq Y \leq a_{h+1}\right\}}| \leq \frac{\gamma_3}{H^{\nu}} \var{\left\{\alpha^{\T}m(Y)\right\}}.
	\end{align*}
\end{definition}

\begin{condition}\label{con5}
$s = |\mathcal{S}| \ll p$ where $\mathcal{S} = \{i|\beta_{j}(i) \neq 0 \quad\text{for some }j, 1\leq j \leq d\}$ and $|\mathcal{S}|$ is the number of elements in the set of $|\mathcal{S}|$.
\end{condition}

\begin{condition}\label{con6}
	Assume $\max_{1<i<p} r_i$ is bounded where $r_i$ is the number of non-zero elements in the $i$th row of $\Sigma$.
\end{condition}

Condition \ref{con1}-\ref{con2} are essential for the unbiasedness of SIR, as discussed in Section \ref{sec:problem-pre}. 
Conditions \ref{con3}-\ref{con5} were introduced by \citet{lin2018consistency} to ensure the consistency of SIR. In particular, Condition \ref{con4} characterizes the smoothness of the central curve by defining its $\nu$-sliced stable property. 

For the ultra-high dimensional setting, Condition \ref{con5} imposes a sparsity structure on the loading's of $\beta$. Furthermore, instead of imposing the ``approximately bandable'' condition on the covariance matrix as utilized in \citep{lin2018consistency}, we only propose the row sparsity structure in Condition \ref{con6}, which is a relative mild assumption. 


 

Now we are ready to state the main theoretical results of FSIR. 
According to algorithms in Section \ref{sec:method}, we first write
\begin{align*}
	\widetilde{M} = \sum_{k=1}^{K}n_k \widetilde{M}^{(k)}/ N = 
	\sum_{k=1}^{K}n_k \widehat{M}^{(k)}/N + \sum_{k=1}^{K}n_k E_0^{(k)}/N
\end{align*}
and $\widehat{{\Lambda}} = \widetilde{M}\widetilde{M}^{\T}$, where $N=\sum_{k=1}^{K}n_k$. 
Theorem \ref{ld-consistency1} provides the convergence rate of kernel matrix $\widehat{{\Lambda}}$.

\begin{theorem}\label{ld-consistency1}
Under conditions \ref{con1}-\ref{con4}, if ${n_{\epsilon,\delta,p,R}^2 \epsilon^2}/({R^4 H^{2\nu-1}p^3 \log{\frac{2}{\delta}}}) \rightarrow \infty$, then we have
\begin{align}\label{thm31}
\|\widehat{\Lambda} - \Lambda_p\|_2 = O_p\left(\frac{1}{H^{\nu+1}} + \frac{H p}{N} + {\frac{ p^{1/2}}{N^{1/2}}}\right).
\end{align}
\end{theorem}


From Section \ref{sec:problem-fed} we know that 
\begin{align*}
\widetilde{\Sigma} = \frac{1}{N} X X^{\T} + \frac{1}{N}\sum_{k=1}^{K} n_k (c^{(k)}_{R})^2 A^{(k)} = \frac{1}{N}X X^{\T} + \sum_{k=1}^{K} (c^{(k)}_{R})^2 A,
\end{align*}
where $A = n_k A^{(k)}/N$ is a symmetric random Gaussian matrix and $A_{i,j} \sim \mathcal{N}(0, \sigma_{k}^{2})$ for $i \geq j$ where
\begin{align*}
\sigma_{x} = \frac{p+1}{N \epsilon_{x}}\left\{2\log{\left(\frac{p^2 +p}{2\sqrt{2\pi}\delta_x} \right)}\right\}^{1/2} + \frac{1}{N{\epsilon_{x}}^{1/2}}.
\end{align*}

Then we have the following Lemma:
\begin{lemma}\label{lem3}
Under condition \ref{con3}, we have 
\begin{align*}
\|\widetilde{\Sigma} - \Sigma\|_2 \rightarrow 0
\end{align*}
as $p/N \rightarrow 0$ and $N \rightarrow \infty$. It is also easy to see that
\begin{align*}
\|\widetilde{\Sigma}^{-1} - {\Sigma^{-1}}\|_2 \rightarrow 0.
\end{align*}
\end{lemma}
With lemma \ref{lem3}, we can show that our FSIR procedure provides a consistent estimate of the sufficient dimension reduction space.

\begin{theorem}\label{ld-consistency2}
Under conditions \ref{con1}-\ref{con4} and assuming that $p/N \rightarrow 0$, we have
\begin{align*}
\|\widetilde{\Sigma}^{-1}\widehat{\Lambda} - \Sigma^{-1}\Lambda_p\|_2 \rightarrow 0
\end{align*}
as $N \rightarrow \infty$ with probability converging to one.
\end{theorem}

As demonstrated by Theorems \ref{ld-consistency1} and \ref{ld-consistency2}, FSIR is capable of obtaining a reliable estimation of the central space in low dimensional scenarios. Additionally, we will illustrate how FSIR can effectively address high dimensional settings through the implementation of a screening procedure. Prior to that, we will outline the key properties of the screening procedure.

Let the sample mean in the $h$th slice be $\bar{x}_h$, and we define the inclusion set and exclusion set below, which depend on a threshold value $t$,
\begin{align*}
\mathcal{T} = \left\{j| {E}\{x(j)|y\} \quad \text{is not a constant} \right\},
\end{align*}
\begin{align*}
\mathcal{I}_h = \left\{ j| |\bar{x}_h (j)| > t \right\}, h = 1,..., H.
\end{align*}
\begin{align*}
\mathcal{I} = \cup_h \mathcal{I}_h = \left\{ j|\text{there exists a} \quad h \in [1,H], \text{s.t.}\quad |\bar{x}_h (j)| > t\right\},
\end{align*}
\begin{align*}
\mathcal{I}^{c} = \left\{j|\text{for all}\quad h \in [1, H], \quad |\bar{x}_h(j)| < t \right\}.
\end{align*}
Furthermore, we write the smallest sample size of slices as $n_{\epsilon,\delta,s,R}$ in high dimension settings.

\begin{assumption}\label{ass1}
	Signal strength: $\exists C > 0$ and $\omega > 0$ such that $\mathrm{var}[E\{x(k)|y\}] > C s^{-\omega}$ when ${E}\{x(k)|y\}$ is not a constant.
\end{assumption}

\begin{proposition}\label{prop:scr}
Under condition \ref{con1} - \ref{con6} and assumption \ref{ass1}, and let $t = Cs^{-{\omega}/{2}}$ for some constant $C > 0$ such that $t < 2 C_{\mu} s^{-{\omega}/{2}}$, we have

i) $\mathcal{T} \subset \mathcal{I}$ holds with probability at least
\begin{align*}
&1 -  C_1 \exp\left\{-C_2 \frac{n_{\epsilon,\delta,s,R} s^{-\omega}}{H\nu^2} + C_3 \log{(H)}+ C_4 \log{(s)}\right\}\\
&- C_5 \exp \left\{\frac{-n_{\epsilon,\delta,s,R} s^{-\omega}}{2 C H \tau ^{2}+2 s^{-\frac{\omega}{2}} \tau} + C_6 \log{(s)} + C_7 \log{(H)}\right\}.
\end{align*}

ii) $\mathcal{T}^c \subset \mathcal{I}^c$ holds with probability at least
\begin{align*}
1 - C_1 \exp \left\{\frac{-n_{\epsilon,\delta,s,R} s^{-\omega}}{2 C H \tau ^{2}+2 s^{-\frac{\omega}{2}} \tau} + C_2 \log{(H)} + C_3 \log{(p - cs)}\right\}.
\end{align*}
\end{proposition}

This Proposition has a simple implication. We may choose \[H = o({C n_{\epsilon,\delta,s,R} s^{-\omega}}{\tau^{-2} /\log{p}})\] so that
\[ \frac{C n_{\epsilon,\delta,s,R} s ^{-\omega}}{H\tau^2} \gg \log{p}. \]

With the help of proposition \ref{prop:scr}, we provide the properties of screening procedure in the theorem below.

\begin{theorem}\label{thm:scr}
Let $\mathcal{T}_0 = \left\{j \mid \operatorname{multiplicity}(j) > \frac{K}{2}\right\}$,
we have

i) $\mathcal{T} \subset \mathcal{T}_0$ holds with probability at least
\begin{align*}
1 - C s K^{3/2} (\frac{K^2 p_1}{e^2})^{K/2},
\end{align*}
where 
\begin{align*}
p_1 =  C_1 \exp\left\{-C_2 \frac{n_{\epsilon.\delta,s,R} s^{-\omega}}{H\nu^2} + C_3 \log{(H)}\right\}
+ C_4 \exp \left\{\frac{-n_{\epsilon,\delta,s,R} s^{-\omega}}{2 C H \tau ^{2}+2 s^{-\frac{\omega}{2}} \tau} + C_5 \log{(H)}\right\}.
\end{align*}

ii) $\mathcal{T}^c \subset \mathcal{T}_0^c$ holds with probability at least
\begin{align*}
1 - C s K^{3/2} (\frac{K^2 p_1}{e^2})^{K/2},
\end{align*}
where 
\begin{align*}
p_2 =  C_1 \exp \left\{\frac{-n_{\epsilon,\delta,s,R} s^{-\omega}}{2 C H \tau ^{2}+2 s^{-\frac{\omega}{2}} \tau} + C_2 \log{(H)}\right\}.
\end{align*}
\end{theorem}

It can be implied that we only need \[H \log{H} \ll \min{({n_{\epsilon,\delta,s,R}}{s^{-\omega} \nu^{-2}}, {n_{\epsilon,\delta,s,R}}{H^{-1}\tau^{-2}s^{-\omega}})}.\] 
Under this assumption, we have $\mathcal{T} = \mathcal{T}_0$ with probability converging to one.
Thus we know $\mathcal{T}_0 = \mathcal{T}$ with probability converging to one, allowing us to draw some conclusions for high-dimensional FSIR which are shown in the following two theorems. 

\begin{theorem}\label{hd-consistency1}
Under condition \ref{con1} - \ref{con6} and assumption \ref{ass1} and choosing the same $t$ as Theorem \ref{thm:scr}, if $H \log{H} \ll \min{({n_{\epsilon,\delta,s,R}}{s^{-\omega} \nu^{-2}}, {n_{\epsilon,\delta,s,R}}{H^{-1}\tau^{-2}s^{-\omega}})}$, we have
\begin{align*}
\|e(\widehat{\Lambda}^{\mathcal{T}_0 , \mathcal{T}_0}) - \Lambda\|_2 \rightarrow 0
\end{align*}
as $n \rightarrow \infty $ with probability converging to one.
\end{theorem}

\begin{theorem}\label{hd-consistency2}
Under the same conditions and assumptions in Theorem \ref{hd-consistency1}
\begin{align*}
\|e\left\{(\widehat{\Sigma}^{\mathcal{T}_0, \mathcal{T}_0})^{-1}\widehat{\Lambda}^{\mathcal{T}_0 , \mathcal{T}_0}\right\} - e\left\{(\Sigma^{\mathcal{T}_0, \mathcal{T}_0})^{-1}\Lambda^{\mathcal{T}_0, \mathcal{T}_0}\right\}\|_2 \rightarrow 0
\end{align*}
as $n \rightarrow \infty$ with probability converging to 1.
\end{theorem}

By the sparsity of $\beta$,  $e\left\{(\Sigma^{\mathcal{T}, \mathcal{T}})^{-1}\Lambda^{\mathcal{T}, \mathcal{T}}\right\}$ can be regarded as the true central subspace, thus it is reasonable using $e\left\{(\widehat{\Sigma}^{\mathcal{T}_0, \mathcal{T}_0})^{-1}\widehat{\Lambda}^{\mathcal{T}_0 , \mathcal{T}_0}\right\}$ as the estimation of central subspace when $\mathcal{T}_0 = \mathcal{T}$.

\section{Simulation studies}\label{sec:simu}
\subsection{Simulation setting}
We consider five models in simulation studies; see details below. 
Among them, Model (I) uses the logistic regression to generates a binary response $Y$ by a single index $\beta_1^{\T} X$, where $\beta_1 \in \mathbb{R}^p$. 
Model (II) is another single index model emphasizing a heterogeneous covariance structure.  
For both of them, the structure dimension $\operatorname{dim}(\mathcal{S}_{Y|X})=1$.
Model (III) and (IV) further consider the double index setting. 
Without loss of generality, we specify a $\beta_2 \in \mathbb{R}^p$ which is orthogonal to $\beta_1$, 
so $\operatorname{dim}(\mathcal{S}_{Y|X})=2$. 
For each of the Model (II) to (IV), assume $\epsilon \sim N(0, 1)$ and $\epsilon \indep X$. 
Model (V) presents an inverse regression model with $\epsilon \sim N(0,I_p)$ and $\epsilon \indep Y$. 
Denote matrix $\Gamma \triangleq (\beta_1,\beta_2) \in \mathbb{R}^{p \times 2}$, then it is semi-orthogonal and $\mathcal{S}_{Y|X}=\operatorname{span}(\Gamma)$; see \citep{cook2008principal}.
\begin{itemize}
	\item (I)  $Y=\idop(1/\{1+\exp(\beta_1^{\T} X)\})$, where $X \sim N(0,I_{p})$.
	\item (II) $Y = \frac{1}{0.5+ (\beta_1^{\T}X+1)^2}+\epsilon$ where $X \sim N(0,\Sigma)$ with $\Sigma_{i,j}=0.5^{|i-j|}$ for $1\leq i,j \leq p$.
	\item (III) $Y =\frac{\beta_1^{\T} X}{(\beta_2^{\T}X)^3 + 1} + \epsilon$ where $X \sim N(0,I_{p})$.
	\item (IV) $Y=\sin(\beta_1^{\T} X)\exp(\beta_2^{\T}X+ \epsilon)$ where $X \sim N(0,\Sigma)$ with $\Sigma_{i,j}=0.5^{|i-j|}$ for $1\leq i,j \leq p$.
	\item (V) $X=\Gamma f(Y)+ \epsilon$ where $\Gamma \triangleq (\beta_1,\beta_2) \in \mathbb{R}^{p \times 2}$, $f(Y)=(Y,Y^2)^{\T}$ and $Y \sim N(0,1)$.	
\end{itemize}

In the low-dimensional setting, we take a fixed covariate dimension $p=10$. For model (I) and (II), generate $\beta_{1,j} \sim \operatorname{Unif} (0.4, 0.8)$ for $1\leq j \leq p$ and set $\beta_1$ by normalizing $(\beta_{1,1},\dots,\beta_{1,p})^{\T}$. 
For model (III) to (V), specify $\beta_1 = (1,1,1,1,1,0,\dots,0)^{\T}/\sqrt{5}$ and $\beta_2 = (0,0,0,0,0,1,\dots,1)^{\T}/\sqrt{5}$. 
In the high-dimensional setting, specify $p = 500, 1000, 2000$, separately. 
Denote the size of active set by $s$ and let $s_0 \triangleq \ceil{s/2}$. When $p \leq 500$, we take $s=5$, otherwise, $s=10$. 
For model (I) and (II), generate $\beta_{1,j} \sim \operatorname{Unif} (0.4, 0.8)$ for $1\leq j \leq s$ and $\beta_{1,j}=0$ for $s < j \leq p$.
For model (III) to (V), set $\beta_{1,j} =1$ for $1\leq j \leq s_0$ and $\beta_{1,j}=0$ for $s_0 < j \leq p$; 
similarly, set $\beta_{2,j} =1$ for $s-s_0 \leq j \leq s$ and $\beta_{2,j}=0$ otherwise. 
Finally, $\beta_1$ and $\beta_2$ are obtained by normalizing $(\beta_{1,1},\dots,\beta_{1,p})^{\T}$ and $(\beta_{2,1},\dots,\beta_{2,p})^{\T}$ respectively.

To evaluate the accuracy of the estimated $\widehat{\beta}$, we use the so-called projection loss $$d(\widehat{\beta},\beta)=||\widehat{\beta}(\widehat{\beta}^{\T}\widehat{\beta})^{-1}\widehat{\beta}^{\T} - \beta(\beta^{\T}\beta)^{-1}\beta^{\T}||_F,$$
which measures the distance between the subspace spanned by $\widehat{\beta}$ and the target subspace spanned by $\beta$ in a coordinate-free manner. 
The angle between $\widehat{\beta}$ and $\beta$, denoted by $\angle(\widehat{\beta},\beta)$, is also presented in a more intuitive way.

\subsection{Results}
\begin{figure}[!h]
	\centering
	\includegraphics[width=\textwidth]{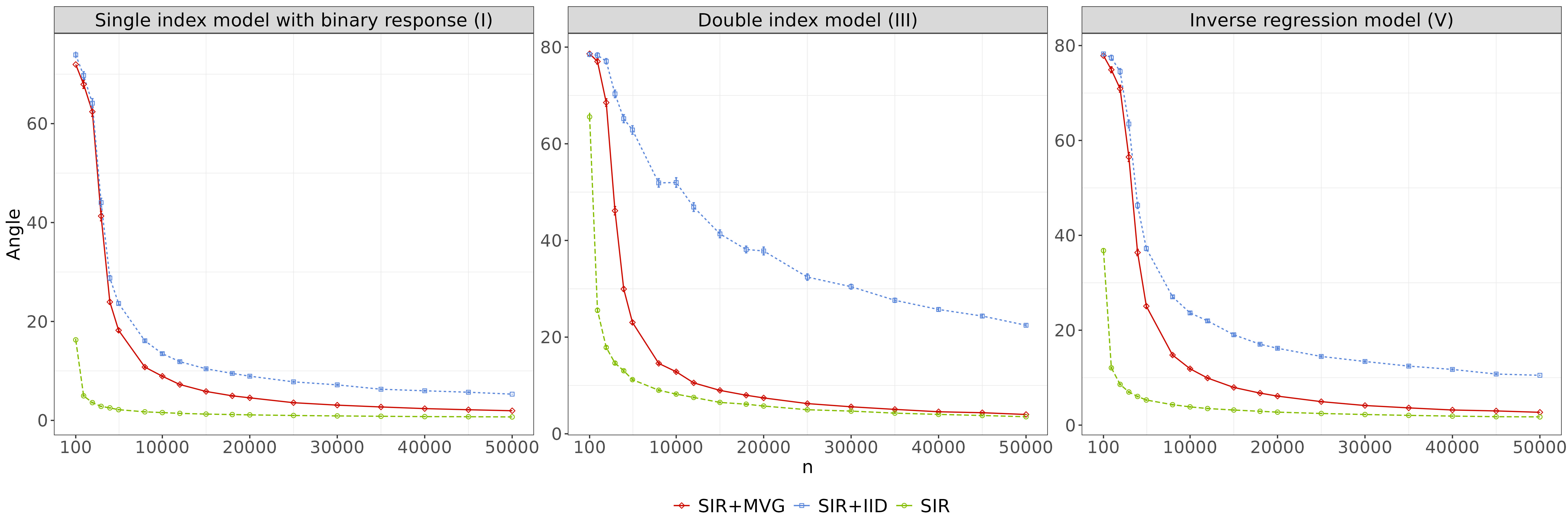}
	\caption{The cost of $(\epsilon, \delta)$-privacy in the SIR estimation when $K=1$.}
	\label{fig:dp-sir}	
\end{figure}

The first part of our numerical study illustrates the cost of $(\epsilon, \delta)$-privacy in SIR estimation under the ordinary circumstance which means $K=1$.
We compare two strategies for ensuring the privacy of the slice mean matrix: the vanilla i.i.d. Gaussian mechanism and the improved vectorized Gaussian mechanism (VGM). 
We denote the SIR combined with the former strategy as SIR-IID and the latter as SIR-VGM. 
To make comparison easier,  we generate datasets only for models (I), (III), and (V) separately, varying the sample size $n$ from $100$ to $50000$. 
We take the privacy level sets $(\epsilon_x, \delta_x)$ at $(1,1/n^{1.1})$ and $(\epsilon_m, \delta_m)$ at $(1,1/n^{1.1})$ separately.
In each experiment, we calculate the angle $\angle(\widehat{\beta},\beta)$, where $\widehat{\beta}$ is estimated using the original SIR, SIR-IID, and SIR-VGM, respectively. Figure \ref{fig:dp-sir} plots the averaged angle obtained from $400$ replications, with an error bar locating $90\%$ of the results.
The plot clearly shows that SIR-VGM, indicated by the red curve, significantly outperforms SIR-IID, represented by the light blue curve, across all three models. Notably, SIR-IID yields unsatisfactory results for Model (III) and (V), both of which assume a structure dimension $d=2$. 
Furthermore, as the sample size increases, the curve of SIR-VGM approaches the lower bound achieved by the original SIR method. This observation suggests that the vectorized Gaussian mechanism introduces only a negligible loss of accuracy when the sample size is sufficiently large.

\begin{figure}[!ht]
	\centering
	\includegraphics[width=\textwidth]{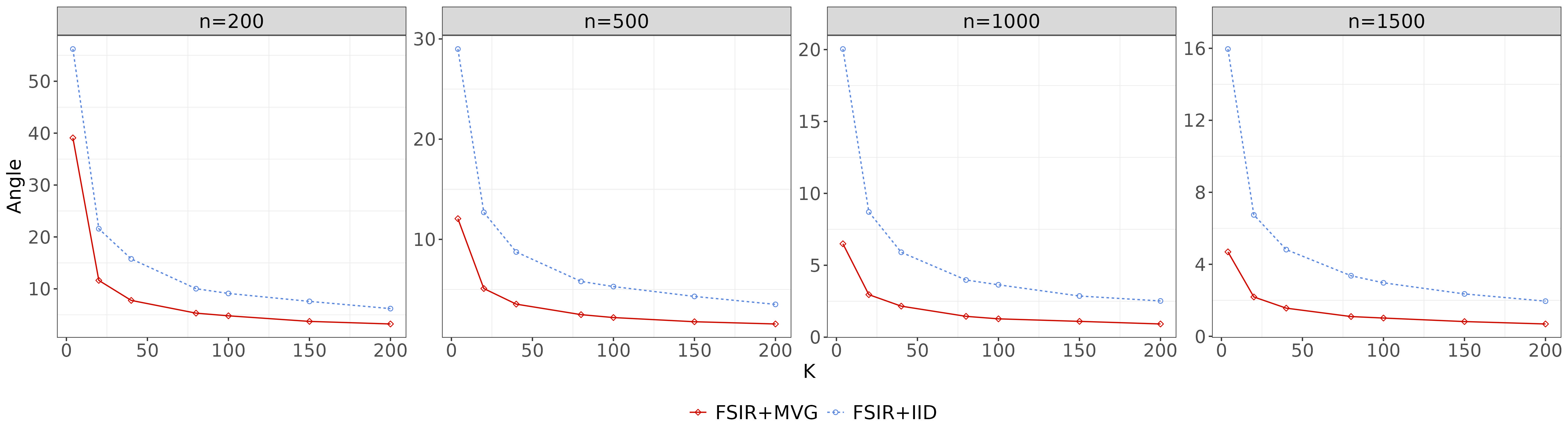}
	\caption{Averaged angle of FSIR-IID and FSIR-VGM calculated for model (I) plotted against client number $K$. From left to right, the client sample size $n$ ranges from $100$ to $1000$.}
	\label{fig:sim1-K}
\end{figure}

When $K>1$, the original SIR method can not be applied. The subsequent study investigates the effectiveness of the federated paradigm ($K>1$) by reporting the performance of two approaches, namely FSIR-IID and FSIR-VGM, under Model (I), against varying $K$. As mentioned earlier, Model (I) involves a binary response, which implies that the estimated $\widehat{\beta}$ is scaled to the discriminant direction LDA gives. Consequently, this study also evaluates the performance of two differentially private LDA methods within the federated framework. 
We take the client sample size $n \in \{100, 500, 1000\}$ and the client number $K$ ranges from $2$ to $200$. 
The privacy parameters are unchanged, except for $\epsilon_x = 3$.
The selection of a relatively large value of $\epsilon_x$ in this context is intended to regulate the scale of the added noise on the covariance matrix when comparing FSIR-IID and FSIR-VGM.
In each setting, we calculate $\angle(\widehat{\beta},\beta)$ and repeat the corresponding experiment $400$ times. 
Figure \ref{fig:sim1-K} depicts the averaged angles for two methods against $K$. 
The results demonstrate that both methods achieve smaller angles as $K$ increases. 
Moreover, FSIR-VGM performs better than FSIR-IID across all settings, particularly when $n$ is large.  

For a comprehensive assessment of FSIR-IID and FSIR-VGM, we evaluate their performance across all five models using the subspace distance metric $d(\beta, \widehat{\beta})$, which is defined above. 
For simplicity, we fix the slice number to $H=8$ when the response is continuous, and assume that $\epsilon_x$ and $\epsilon_m$ share the same value $\epsilon \in \{1,2\}$. 
In the low-dimensional setting, we set $p=10$ and conduct experiments for combinations of client sample size $n \in \{500, 1000, 2500, 5000\}$ and client number $K \in \{1,10, 50, 100\}$.  Table \ref{table:lowdim-1} and \ref{table:lowdim-2} report the averaged $d(\beta, \widehat{\beta})$ based on $400$ replications, with standard errors in parenthesis. Notice when $K=1$, indicating a single client in the federated framework, FSIR-IID (FSIR-VGM) reduces to SIR-IID (SIR-VGM). 
For high-dimensional cases, we fix $n=1000$ for each client and set $p=500,1000,2000$. The simulation results are summarized in Table \ref{table:highdim-1} and \ref{table:highdim-2}.
We can see as client number $K$ increases, FSIR-IID and FSIR-VGM get better results, while FSIR-VGM performs significantly better than FSIR-IID in almost all settings.
\begin{table}[!ht]
	\caption{$p=10,\ H=8,\ \epsilon=1$.}
	\tiny
	\begin{tabular}{crrrrrrrrr}
\hline
& \multirow{2}{*}{\textbf{n}} & \multicolumn{4}{c}{\textbf{FSIR-IID}}                                                                                                         & \multicolumn{4}{c}{\textbf{FSIR-VGM}}                                                                                                         \\ \cline{3-10} 
&                             & \multicolumn{1}{c}{\textbf{K=1}} & \multicolumn{1}{c}{\textbf{K=10}} & \multicolumn{1}{c}{\textbf{K=50}} & \multicolumn{1}{c}{\textbf{K=100}} & \multicolumn{1}{c}{\textbf{K=1}} & \multicolumn{1}{c}{\textbf{K=10}} & \multicolumn{1}{c}{\textbf{K=50}} & \multicolumn{1}{c}{\textbf{K=100}} \\ \hline
\multirow{4}{*}{I}   & 500                         & 1.306 (0.13)                     & 1.268 (0.15)                      & 0.650 (0.19)                      & 0.382 (0.10)                       & 1.290 (0.14)                     & 1.298 (0.12)                      & 0.598 (0.17)                      & 0.384 (0.09)                       \\
& 1000                        & 1.260 (0.15)                     & 0.547 (0.13)                      & 0.206 (0.05)                      & 0.141 (0.03)                       & 1.277 (0.16)                     & 0.500 (0.15)                      & 0.190 (0.04)                      & 0.128 (0.03)                       \\
& 2500                        & 1.131 (0.23)                     & 0.327 (0.07)                      & 0.135 (0.03)                      & 0.091 (0.02)                       & 1.092 (0.25)                     & 0.287 (0.08)                      & 0.115 (0.03)                      & 0.078 (0.02)                       \\
& 5000                        & 0.563 (0.13)                     & 0.186 (0.04)                      & 0.079 (0.02)                      & 0.054 (0.01)                       & 0.428 (0.11)                     & 0.140 (0.03)                      & 0.058 (0.01)                      & 0.040 (0.01)                       \\
&                             &                                  &                                   &                                   &                                    &                                  &                                   &                                   &                                    \\
\multirow{4}{*}{II}  & 500                         & 1.360 (0.07)                     & 1.374 (0.06)                      & 1.292 (0.14)                      & 1.061 (0.22)                       & 1.328 (0.10)                     & 1.275 (0.14)                      & 1.167 (0.23)                      & 0.760 (0.26)                       \\
& 1000                        & 1.370 (0.06)                     & 1.311 (0.11)                      & 0.788 (0.18)                      & 0.565 (0.14)                       & 1.274 (0.16)                     & 1.057 (0.27)                      & 0.369 (0.09)                      & 0.241 (0.07)                       \\
& 2500                        & 1.381 (0.05)                     & 1.107 (0.15)                      & 0.614 (0.15)                      & 0.445 (0.10)                       & 1.263 (0.16)                     & 0.547 (0.18)                      & 0.230 (0.06)                      & 0.163 (0.04)                       \\
& 5000                        & 1.331 (0.10)                     & 0.854 (0.16)                      & 0.438 (0.11)                      & 0.308 (0.08)                       & 0.938 (0.28)                     & 0.281 (0.08)                      & 0.136 (0.03)                      & 0.111 (0.02)                       \\
&                             &                                  &                                   &                                   &                                    &                                  &                                   &                                   &                                    \\
\multirow{4}{*}{III} & 500                         & 1.780 (0.13)                     & 1.760 (0.13)                      & 1.364 (0.18)                      & 1.156 (0.23)                       & 1.773 (0.13)                     & 1.716 (0.14)                      & 1.303 (0.18)                      & 0.952 (0.25)                       \\
& 1000                        & 1.750 (0.13)                     & 1.353 (0.18)                      & 0.846 (0.23)                      & 0.561 (0.16)                       & 1.648 (0.20)                     & 0.754 (0.14)                      & 0.280 (0.05)                      & 0.193 (0.04)                       \\
& 2500                        & 1.628 (0.18)                     & 1.142 (0.22)                      & 0.577 (0.18)                      & 0.387 (0.09)                       & 1.365 (0.25)                     & 0.422 (0.08)                      & 0.169 (0.03)                      & 0.113 (0.02)                       \\
& 5000                        & 1.347 (0.17)                     & 0.835 (0.25)                      & 0.355 (0.09)                      & 0.237 (0.06)                       & 0.641 (0.13)                     & 0.215 (0.04)                      & 0.087 (0.02)                      & 0.062 (0.01)                       \\
&                             &                                  &                                   &                                   &                                    &                                  &                                   &                                   &                                    \\
\multirow{4}{*}{IV}  & 500                         & 1.762 (0.13)                     & 1.704 (0.16)                      & 1.452 (0.25)                      & 1.089 (0.24)                       & 1.715 (0.17)                     & 1.679 (0.17)                      & 1.431 (0.21)                      & 0.957 (0.25)                       \\
& 1000                        & 1.778 (0.13)                     & 1.427 (0.22)                      & 0.577 (0.11)                      & 0.401 (0.07)                       & 1.698 (0.16)                     & 1.241 (0.27)                      & 0.459 (0.08)                      & 0.314 (0.06)                       \\
& 2500                        & 1.724 (0.16)                     & 0.974 (0.18)                      & 0.421 (0.08)                      & 0.302 (0.06)                       & 1.579 (0.22)                     & 0.735 (0.18)                      & 0.287 (0.06)                      & 0.211 (0.04)                       \\
& 5000                        & 1.481 (0.21)                     & 0.613 (0.11)                      & 0.283 (0.05)                      & 0.208 (0.04)                       & 1.174 (0.26)                     & 0.361 (0.07)                      & 0.178 (0.03)                      & 0.148 (0.02)                       \\
&                             &                                  &                                   &                                   &                                    &                                  &                                   &                                   &                                    \\
\multirow{4}{*}{V}   & 500                         & 1.774 (0.12)                     & 1.722 (0.16)                      & 1.054 (0.24)                      & 0.629 (0.12)                       & 1.702 (0.16)                     & 1.683 (0.19)                      & 1.034 (0.22)                      & 0.580 (0.12)                       \\
& 1000                        & 1.741 (0.15)                     & 0.952 (0.16)                      & 0.367 (0.06)                      & 0.245 (0.05)                       & 1.669 (0.18)                     & 0.817 (0.17)                      & 0.296 (0.05)                      & 0.194 (0.04)                       \\
& 2500                        & 1.630 (0.17)                     & 0.604 (0.11)                      & 0.251 (0.05)                      & 0.173 (0.03)                       & 1.497 (0.23)                     & 0.462 (0.08)                      & 0.181 (0.03)                      & 0.124 (0.02)                       \\
& 5000                        & 1.005 (0.17)                     & 0.364 (0.06)                      & 0.154 (0.03)                      & 0.107 (0.02)                       & 0.739 (0.15)                     & 0.226 (0.04)                      & 0.092 (0.02)                      & 0.063 (0.01)                       \\ \hline
\end{tabular}
	\label{table:lowdim-1}
\end{table}

\begin{table}[!ht]
	\caption{$p=10,\ H=8,\ \epsilon=2$.}
	\tiny
	\begin{tabular}{ccrrrrrrrr}
		\hline
		& \multirow{2}{*}{\textbf{n}} & \multicolumn{4}{c}{\textbf{FSIR-IID}}                                                                                                         & \multicolumn{4}{c}{\textbf{FSIR-VGM}}                                                                                                         \\ \cline{3-10} 
		&                             & \multicolumn{1}{c}{\textbf{K=1}} & \multicolumn{1}{c}{\textbf{K=10}} & \multicolumn{1}{c}{\textbf{K=50}} & \multicolumn{1}{c}{\textbf{K=100}} & \multicolumn{1}{c}{\textbf{K=1}} & \multicolumn{1}{c}{\textbf{K=10}} & \multicolumn{1}{c}{\textbf{K=50}} & \multicolumn{1}{c}{\textbf{K=100}} \\ \hline
		\multirow{4}{*}{I}   & 500                         & 1.294 (0.13)                     & 0.890 (0.29)                      & 0.292 (0.07)                      & 0.190 (0.05)                       & 1.279 (0.14)                     & 0.859 (0.26)                      & 0.278 (0.07)                      & 0.181 (0.04)                       \\
		& 1000                        & 0.939 (0.24)                     & 0.277 (0.06)                      & 0.111 (0.02)                      & 0.077 (0.02)                       & 0.850 (0.26)                     & 0.234 (0.06)                      & 0.094 (0.02)                      & 0.064 (0.01)                       \\
		& 2500                        & 0.527 (0.12)                     & 0.190 (0.04)                      & 0.079 (0.02)                      & 0.054 (0.01)                       & 0.407 (0.11)                     & 0.143 (0.04)                      & 0.060 (0.01)                      & 0.041 (0.01)                       \\
		& 5000                        & 0.323 (0.08)                     & 0.113 (0.03)                      & 0.047 (0.01)                      & 0.034 (0.01)                       & 0.216 (0.05)                     & 0.076 (0.02)                      & 0.032 (0.01)                      & 0.022 (0.01)                       \\
		&                             &                                  &                                   &                                   &                                    &                                  &                                   &                                   &                                    \\
		\multirow{4}{*}{II}  & 500                         & 1.370 (0.07)                     & 1.366 (0.07)                      & 0.926 (0.18)                      & 0.653 (0.16)                       & 1.302 (0.13)                     & 1.204 (0.20)                      & 0.557 (0.16)                      & 0.373 (0.13)                       \\
		& 1000                        & 1.381 (0.05)                     & 1.037 (0.18)                      & 0.549 (0.13)                      & 0.401 (0.10)                       & 1.208 (0.20)                     & 0.448 (0.13)                      & 0.205 (0.05)                      & 0.146 (0.04)                       \\
		& 2500                        & 1.325 (0.10)                     & 0.869 (0.17)                      & 0.426 (0.11)                      & 0.296 (0.08)                       & 0.950 (0.23)                     & 0.280 (0.07)                      & 0.141 (0.03)                      & 0.116 (0.02)                       \\
		& 5000                        & 1.153 (0.15)                     & 0.661 (0.16)                      & 0.316 (0.08)                      & 0.220 (0.06)                       & 0.488 (0.13)                     & 0.174 (0.04)                      & 0.106 (0.02)                      & 0.094 (0.02)                       \\
		&                             &                                  &                                   &                                   &                                    &                                  &                                   &                                   &                                    \\
		\multirow{4}{*}{III} & 500                         & 1.768 (0.13)                     & 1.540 (0.15)                      & 1.065 (0.25)                      & 0.713 (0.20)                       & 1.724 (0.14)                     & 1.296 (0.24)                      & 0.453 (0.09)                      & 0.291 (0.05)                       \\
		& 1000                        & 1.558 (0.16)                     & 1.062 (0.24)                      & 0.489 (0.12)                      & 0.323 (0.08)                       & 1.147 (0.23)                     & 0.365 (0.07)                      & 0.150 (0.03)                      & 0.105 (0.02)                       \\
		& 2500                        & 1.355 (0.17)                     & 0.823 (0.23)                      & 0.345 (0.09)                      & 0.230 (0.05)                       & 0.676 (0.12)                     & 0.226 (0.04)                      & 0.098 (0.02)                      & 0.067 (0.01)                       \\
		& 5000                        & 1.133 (0.21)                     & 0.527 (0.12)                      & 0.218 (0.05)                      & 0.149 (0.03)                       & 0.405 (0.08)                     & 0.135 (0.03)                      & 0.058 (0.01)                      & 0.039 (0.01)                       \\
		&                             &                                  &                                   &                                   &                                    &                                  &                                   &                                   &                                    \\
		\multirow{4}{*}{IV}  & 500                         & 1.755 (0.15)                     & 1.626 (0.19)                      & 0.799 (0.18)                      & 0.511 (0.12)                       & 1.727 (0.15)                     & 1.548 (0.20)                      & 0.683 (0.16)                      & 0.440 (0.09)                       \\
		& 1000                        & 1.722 (0.14)                     & 0.835 (0.18)                      & 0.356 (0.07)                      & 0.266 (0.05)                       & 1.533 (0.22)                     & 0.567 (0.12)                      & 0.250 (0.05)                      & 0.193 (0.04)                       \\
		& 2500                        & 1.467 (0.21)                     & 0.612 (0.12)                      & 0.284 (0.06)                      & 0.205 (0.04)                       & 1.115 (0.25)                     & 0.356 (0.07)                      & 0.176 (0.03)                      & 0.146 (0.02)                       \\
		& 5000                        & 1.039 (0.14)                     & 0.425 (0.08)                      & 0.211 (0.04)                      & 0.164 (0.02)                       & 0.521 (0.12)                     & 0.211 (0.04)                      & 0.136 (0.02)                      & 0.125 (0.01)                       \\
		&                             &                                  &                                   &                                   &                                    &                                  &                                   &                                   &                                    \\
		\multirow{4}{*}{V}   & 500                         & 1.757 (0.14)                     & 1.370 (0.19)                      & 0.488 (0.10)                      & 0.322 (0.06)                       & 1.717 (0.15)                     & 1.360 (0.25)                      & 0.441 (0.08)                      & 0.283 (0.05)                       \\
		& 1000                        & 1.467 (0.20)                     & 0.528 (0.10)                      & 0.210 (0.04)                      & 0.144 (0.03)                       & 1.276 (0.25)                     & 0.373 (0.07)                      & 0.149 (0.03)                      & 0.098 (0.02)                       \\
		& 2500                        & 1.008 (0.16)                     & 0.369 (0.07)                      & 0.152 (0.03)                      & 0.103 (0.02)                       & 0.704 (0.15)                     & 0.224 (0.04)                      & 0.091 (0.02)                      & 0.063 (0.01)                       \\
		& 5000                        & 0.677 (0.12)                     & 0.238 (0.04)                      & 0.100 (0.02)                      & 0.069 (0.01)                       & 0.348 (0.06)                     & 0.114 (0.02)                      & 0.048 (0.01)                      & 0.034 (0.01)                       \\ \hline
		\end{tabular}
	\label{table:lowdim-2}
\end{table}

\begin{table}[!ht]
	\tiny
	\caption{$n=1000,\ H=8,\ \epsilon=1$.}
	\begin{tabular}{ccllllllll}
		\hline
		& \multirow{2}{*}{\textbf{p}} & \multicolumn{4}{c}{\textbf{FSIR-IID}}                                                                                                         & \multicolumn{4}{c}{\textbf{FSIR-VGM}}                                                                                                         \\ \cline{3-10} 
		&                             & \multicolumn{1}{c}{\textbf{K=1}} & \multicolumn{1}{c}{\textbf{K=10}} & \multicolumn{1}{c}{\textbf{K=50}} & \multicolumn{1}{c}{\textbf{K=100}} & \multicolumn{1}{c}{\textbf{K=1}} & \multicolumn{1}{c}{\textbf{K=10}} & \multicolumn{1}{c}{\textbf{K=50}} & \multicolumn{1}{c}{\textbf{K=100}} \\ \hline
		I   & 500                         & 0.523 (0.20)                     & 0.192 (0.07)                      & 0.073 (0.02)                      & 0.049 (0.02)                       & 0.308 (0.12)                     & 0.106 (0.05)                      & 0.051 (0.02)                      & 0.033 (0.01)                       \\
		& 1000                        & 1.301 (0.13)                     & 0.914 (0.27)                      & 0.301 (0.08)                      & 0.192 (0.05)                       & 1.281 (0.13)                     & 0.915 (0.27)                      & 0.266 (0.07)                      & 0.185 (0.05)                       \\
		& 2000                        & 1.318 (0.12)                     & 0.926 (0.28)                      & 0.298 (0.08)                      & 0.197 (0.05)                       & 1.302 (0.13)                     & 0.894 (0.26)                      & 0.291 (0.07)                      & 0.191 (0.04)                       \\
		&                             &                                  &                                   &                                   &                                    &                                  &                                   &                                   &                                    \\
		II  & 500                         & 1.387 (0.04)                     & 0.734 (0.22)                      & 0.355 (0.12)                      & 0.245 (0.08)                       & 1.375 (0.05)                     & 0.389 (0.14)                      & 0.198 (0.06)                      & 0.176 (0.05)                       \\
		& 1000                        & 1.400 (0.02)                     & 1.314 (0.11)                      & 0.766 (0.18)                      & 0.482 (0.13)                       & 1.399 (0.02)                     & 1.170 (0.24)                      & 0.552 (0.17)                      & 0.344 (0.10)                       \\
		& 2000                        & 1.402 (0.02)                     & 1.294 (0.14)                      & 0.752 (0.17)                      & 0.516 (0.12)                       & 1.399 (0.02)                     & 1.170 (0.22)                      & 0.565 (0.18)                      & 0.354 (0.10)                       \\
		&                             &                                  &                                   &                                   &                                    &                                  &                                   &                                   &                                    \\
		III & 500                         & 1.917 (0.06)                     & 0.516 (0.21)                      & 0.211 (0.08)                      & 0.132 (0.05)                       & 1.908 (0.06)                     & 0.205 (0.07)                      & 0.079 (0.04)                      & 0.049 (0.02)                       \\
		& 1000                        & 1.961 (0.03)                     & 1.277 (0.26)                      & 0.565 (0.10)                      & 0.424 (0.08)                       & 1.958 (0.03)                     & 1.122 (0.25)                      & 0.464 (0.07)                      & 0.353 (0.07)                       \\
		& 2000                        & 1.962 (0.03)                     & 1.206 (0.21)                      & 0.611 (0.11)                      & 0.501 (0.12)                       & 1.956 (0.03)                     & 1.008 (0.22)                      & 0.529 (0.10)                      & 0.450 (0.12)                       \\
		&                             &                                  &                                   &                                   &                                    &                                  &                                   &                                   &                                    \\
		IV  & 500                         & 1.912 (0.06)                     & 0.888 (0.24)                      & 0.407 (0.09)                      & 0.311 (0.06)                       & 1.908 (0.06)                     & 0.692 (0.23)                      & 0.313 (0.07)                      & 0.260 (0.05)                       \\
		& 1000                        & 1.961 (0.03)                     & 1.728 (0.14)                      & 1.297 (0.27)                      & 0.801 (0.20)                       & 1.954 (0.03)                     & 1.680 (0.14)                      & 1.228 (0.27)                      & 0.762 (0.19)                       \\
		& 2000                        & 1.954 (0.03)                     & 1.731 (0.15)                      & 1.232 (0.26)                      & 0.755 (0.18)                       & 1.950 (0.04)                     & 1.699 (0.17)                      & 1.112 (0.29)                      & 0.702 (0.17)                       \\
		&                             &                                  &                                   &                                   &                                    &                                  &                                   &                                   &                                    \\
		V   & 500                         & 1.912 (0.06)                     & 0.340 (0.10)                      & 0.146 (0.04)                      & 0.090 (0.03)                       & 1.902 (0.07)                     & 0.201 (0.06)                      & 0.080 (0.02)                      & 0.054 (0.02)                       \\
		& 1000                        & 1.957 (0.03)                     & 1.444 (0.20)                      & 0.510 (0.10)                      & 0.334 (0.06)                       & 1.954 (0.03)                     & 1.333 (0.25)                      & 0.453 (0.08)                      & 0.294 (0.05)                       \\
		& 2000                        & 1.958 (0.02)                     & 1.402 (0.22)                      & 0.509 (0.09)                      & 0.340 (0.06)                       & 1.950 (0.03)                     & 1.315 (0.26)                      & 0.470 (0.08)                      & 0.303 (0.05)                       \\ \hline
		\end{tabular}
	\label{table:highdim-1}
\end{table}

\begin{table}[!ht]
	\tiny
	\caption{$n=1000,\ H=8,\ \epsilon=2$.}
	\begin{tabular}{ccllllllll}
		\hline
		& \multirow{2}{*}{\textbf{p}} & \multicolumn{4}{c}{\textbf{FSIR-IID}}                                                                                                         & \multicolumn{4}{c}{\textbf{FSIR-VGM}}                                                                                                         \\ \cline{3-10} 
		&                             & \multicolumn{1}{c}{\textbf{K=1}} & \multicolumn{1}{c}{\textbf{K=10}} & \multicolumn{1}{c}{\textbf{K=50}} & \multicolumn{1}{c}{\textbf{K=100}} & \multicolumn{1}{c}{\textbf{K=1}} & \multicolumn{1}{c}{\textbf{K=10}} & \multicolumn{1}{c}{\textbf{K=50}} & \multicolumn{1}{c}{\textbf{K=100}} \\ \hline
		I   & 500                         & 0.362 (0.13)                     & 0.123 (0.05)                      & 0.047 (0.01)                      & 0.037 (0.01)                       & 0.167 (0.06)                     & 0.069 (0.03)                      & 0.029 (0.01)                      & 0.025 (0.01)                       \\
		& 1000                        & 1.228 (0.17)                     & 0.375 (0.09)                      & 0.161 (0.04)                      & 0.108 (0.02)                       & 1.206 (0.20)                     & 0.348 (0.09)                      & 0.141 (0.03)                      & 0.096 (0.02)                       \\
		& 2000                        & 1.209 (0.19)                     & 0.386 (0.09)                      & 0.160 (0.04)                      & 0.109 (0.02)                       & 1.204 (0.19)                     & 0.337 (0.09)                      & 0.138 (0.03)                      & 0.094 (0.02)                       \\
		&                             &                                  &                                   &                                   &                                    &                                  &                                   &                                   &                                    \\
		II  & 500                         & 1.386 (0.04)                     & 0.511 (0.17)                      & 0.255 (0.08)                      & 0.199 (0.06)                       & 1.370 (0.05)                     & 0.220 (0.07)                      & 0.161 (0.04)                      & 0.156 (0.03)                       \\
		& 1000                        & 1.397 (0.03)                     & 1.063 (0.22)                      & 0.474 (0.13)                      & 0.323 (0.09)                       & 1.396 (0.03)                     & 0.760 (0.27)                      & 0.277 (0.08)                      & 0.183 (0.05)                       \\
		& 2000                        & 1.401 (0.02)                     & 1.003 (0.21)                      & 0.467 (0.12)                      & 0.328 (0.09)                       & 1.394 (0.03)                     & 0.692 (0.22)                      & 0.270 (0.08)                      & 0.190 (0.05)                       \\
		&                             &                                  &                                   &                                   &                                    &                                  &                                   &                                   &                                    \\
		III & 500                         & 1.913 (0.06)                     & 0.317 (0.10)                      & 0.127 (0.05)                      & 0.087 (0.05)                       & 1.906 (0.06)                     & 0.113 (0.06)                      & 0.045 (0.03)                      & 0.035 (0.04)                       \\
		& 1000                        & 1.958 (0.03)                     & 0.776 (0.14)                      & 0.393 (0.09)                      & 0.314 (0.11)                       & 1.956 (0.03)                     & 0.525 (0.08)                      & 0.305 (0.11)                      & 0.265 (0.13)                       \\
		& 2000                        & 1.963 (0.03)                     & 0.813 (0.15)                      & 0.489 (0.11)                      & 0.425 (0.14)                       & 1.957 (0.02)                     & 0.582 (0.10)                      & 0.425 (0.14)                      & 0.394 (0.15)                       \\
		&                             &                                  &                                   &                                   &                                    &                                  &                                   &                                   &                                    \\
		IV  & 500                         & 1.913 (0.06)                     & 0.556 (0.14)                      & 0.288 (0.06)                      & 0.247 (0.04)                       & 1.908 (0.06)                     & 0.372 (0.09)                      & 0.240 (0.04)                      & 0.226 (0.03)                       \\
		& 1000                        & 1.957 (0.03)                     & 1.517 (0.23)                      & 0.633 (0.14)                      & 0.428 (0.09)                       & 1.951 (0.03)                     & 1.431 (0.25)                      & 0.544 (0.12)                      & 0.382 (0.07)                       \\
		& 2000                        & 1.954 (0.03)                     & 1.434 (0.23)                      & 0.586 (0.13)                      & 0.409 (0.09)                       & 1.954 (0.03)                     & 1.389 (0.25)                      & 0.505 (0.11)                      & 0.345 (0.08)                       \\
		&                             &                                  &                                   &                                   &                                    &                                  &                                   &                                   &                                    \\
		V   & 500                         & 1.903 (0.07)                     & 0.230 (0.07)                      & 0.089 (0.03)                      & 0.062 (0.02)                       & 1.901 (0.07)                     & 0.103 (0.03)                      & 0.041 (0.01)                      & 0.027 (0.01)                       \\
		& 1000                        & 1.952 (0.03)                     & 0.697 (0.13)                      & 0.274 (0.05)                      & 0.181 (0.03)                       & 1.953 (0.03)                     & 0.577 (0.12)                      & 0.225 (0.04)                      & 0.150 (0.02)                       \\
		& 2000                        & 1.955 (0.03)                     & 0.708 (0.13)                      & 0.275 (0.05)                      & 0.184 (0.03)                       & 1.953 (0.03)                     & 0.578 (0.10)                      & 0.219 (0.04)                      & 0.151 (0.03)                       \\ \hline
		\end{tabular}
	\label{table:highdim-2}
\end{table}

\section{Real data analysis}\label{sec:real}
\subsection{Human Activity Recognition with Smartphones}
To evaluate our method on classification problem, we examine the human activity recognition dataset \citep{anguita2013public}, which can be downloaded from the UCI website. 
The data were collected from $30$ volunteers who performed six activities (walking, walking upstairs, walking downstairs, sitting, standing, and laying) while wearing a smartphone on their waist. Sensor records were then captured using the smartphone's embedded accelerometer and gyroscope at a rate of $50$ Hz, resulting in a dataset with $10299$ samples and $561$ features. 
Given the sensitivity of cellphone data in real life, we suggest to partition the dataset into $30$ disjoint subsets according to different subjects (regarded as clients), each containing no more than 410 samples, thus representing a high-dimensional setting on any single client.

\begin{figure}[!ht]
	\centering
	\subcaptionbox{FSIR}{\includegraphics[width=0.32\textwidth]{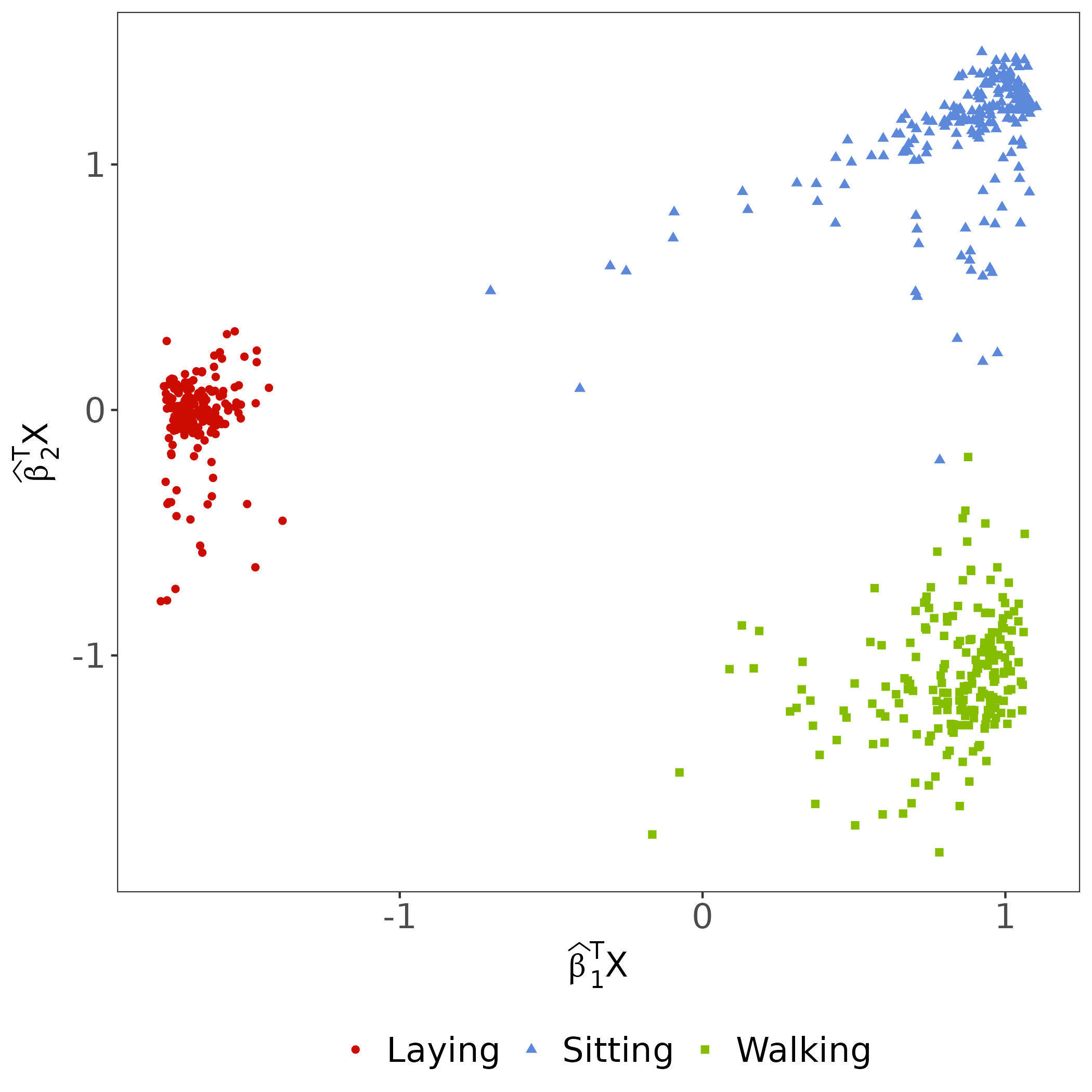}}
	\subcaptionbox{FSIR-IID}{\includegraphics[width=0.32\textwidth]{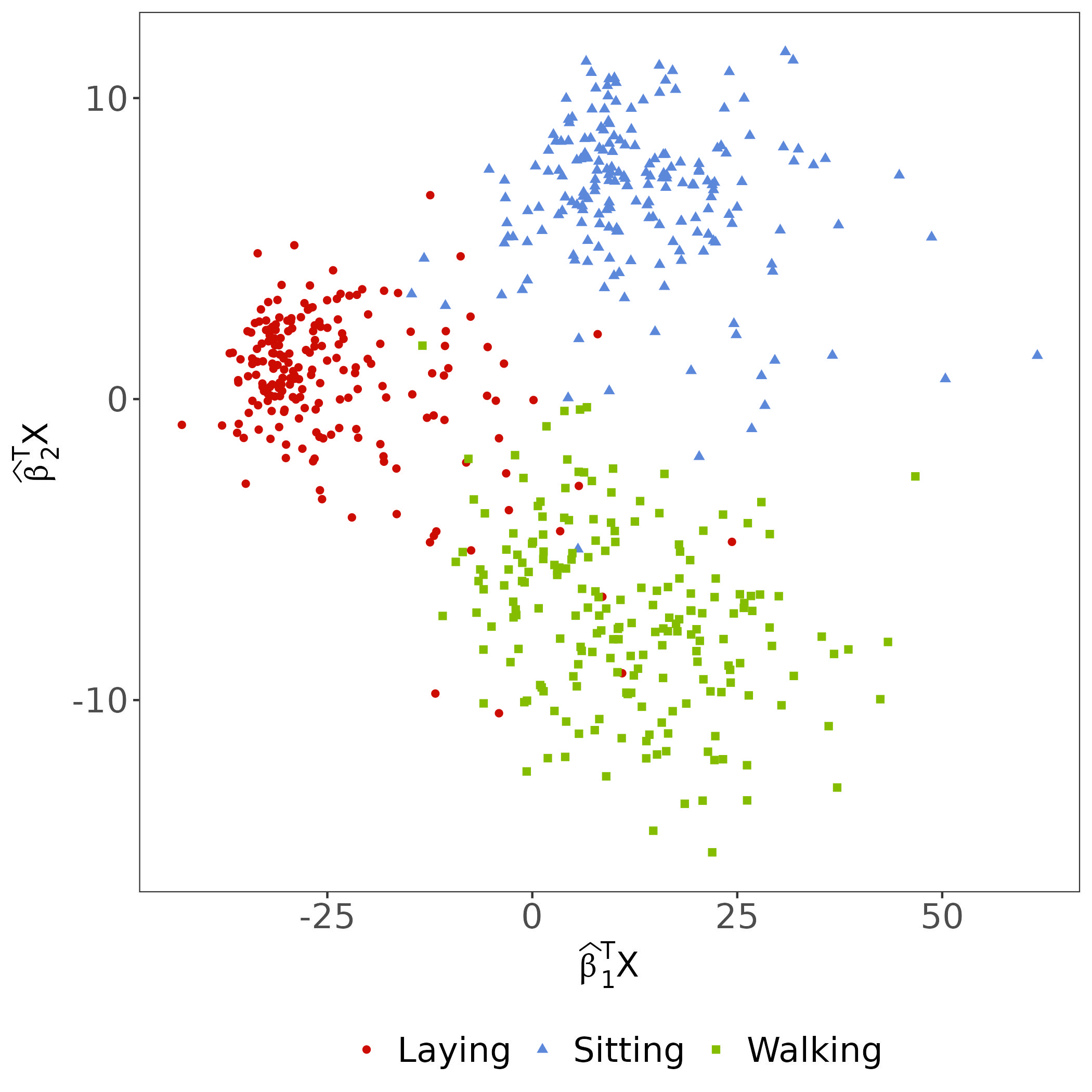}}
	\subcaptionbox{FSIR-VGM}{\includegraphics[width=0.32\textwidth]{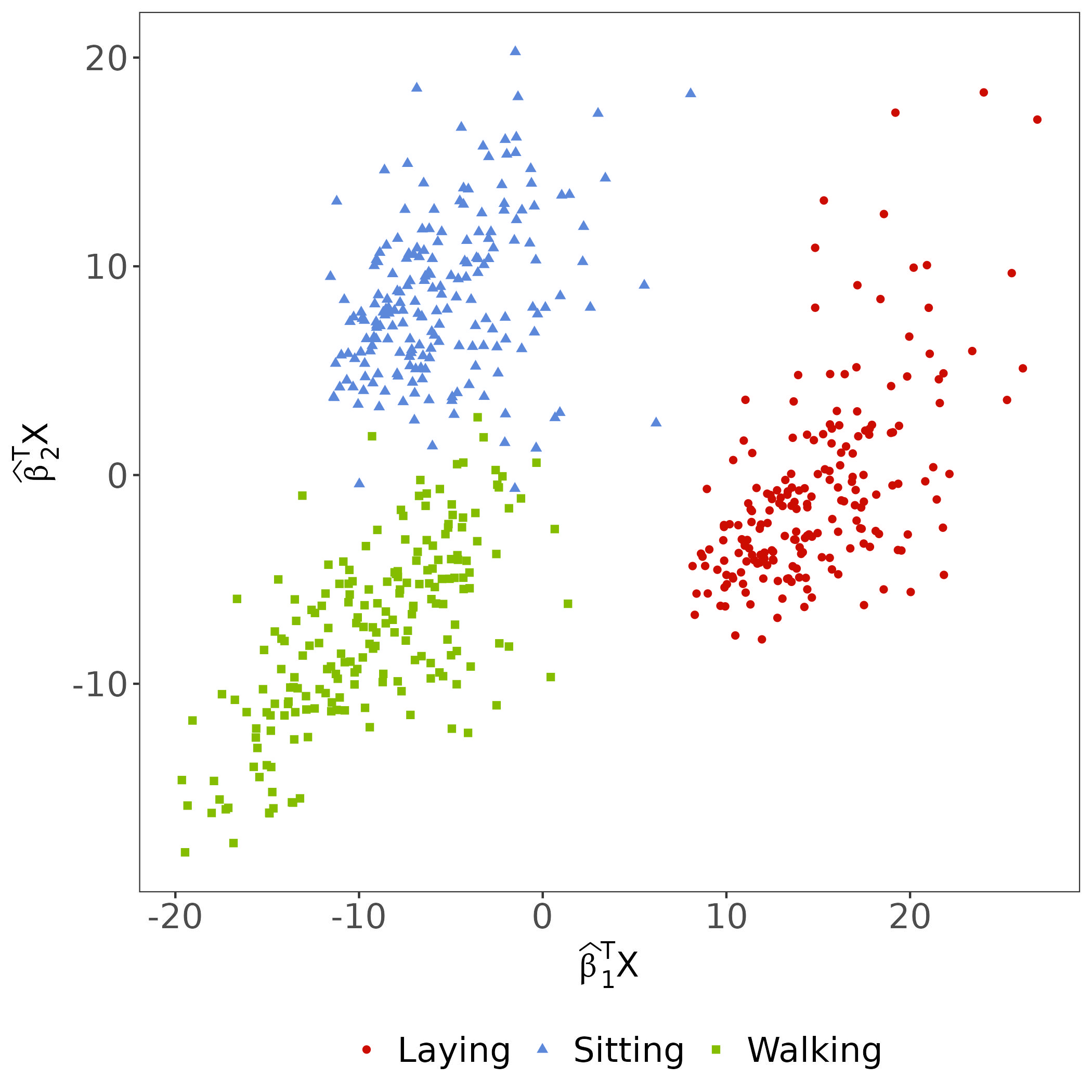}}
	\caption{The scatter plot of the SDR subspace from FSIR, FSIR-IID and FSIR-VGM on a demo subset of the human activity recognition dataset.}
	\label{fig:real-phone}
\end{figure}

In this experiment, to clarify our idea more clearly,  we study three physical activity levels of a person: active (includes walking and walking downstairs/upstairs),  sedentary (includes standing and sitting), and lying down. 
We perform FSIR and its differentially private counterparts on the decentralized dataset. 
With the estimated sparsity parameter $\hat{s}=5$ and structure dimension $\hat{d}=2$, FSIR selects $10$ active covariates to construct a $2$-dimensional SDR subspace $\widehat{\mathcal{S}}_{Y|X}$. Figure \ref{fig:real-phone} depicts $200$ randomly selected samples for each activity on the estimated $\widehat{\mathcal{S}}_{Y|X}$.  
The three activity levels are easily distinguished in each plot, demonstrating the effectiveness of our proposed methods. 
Notably, both FSIR-IID and FSIR-VGM guarantee a $(2,1/n^{1.1})$-level differential privacy, where $n$ is the minimum sample size of a particular activity performed by a single individual across all activities of all subjects. 
We can see when an adequate number of samples is available, FSIR-VGM is preferable. 

\subsection{Airline on-time performance}
\begin{figure}[!ht]
	\centering
	\subcaptionbox{FSIR}{\includegraphics[width=0.32\textwidth]{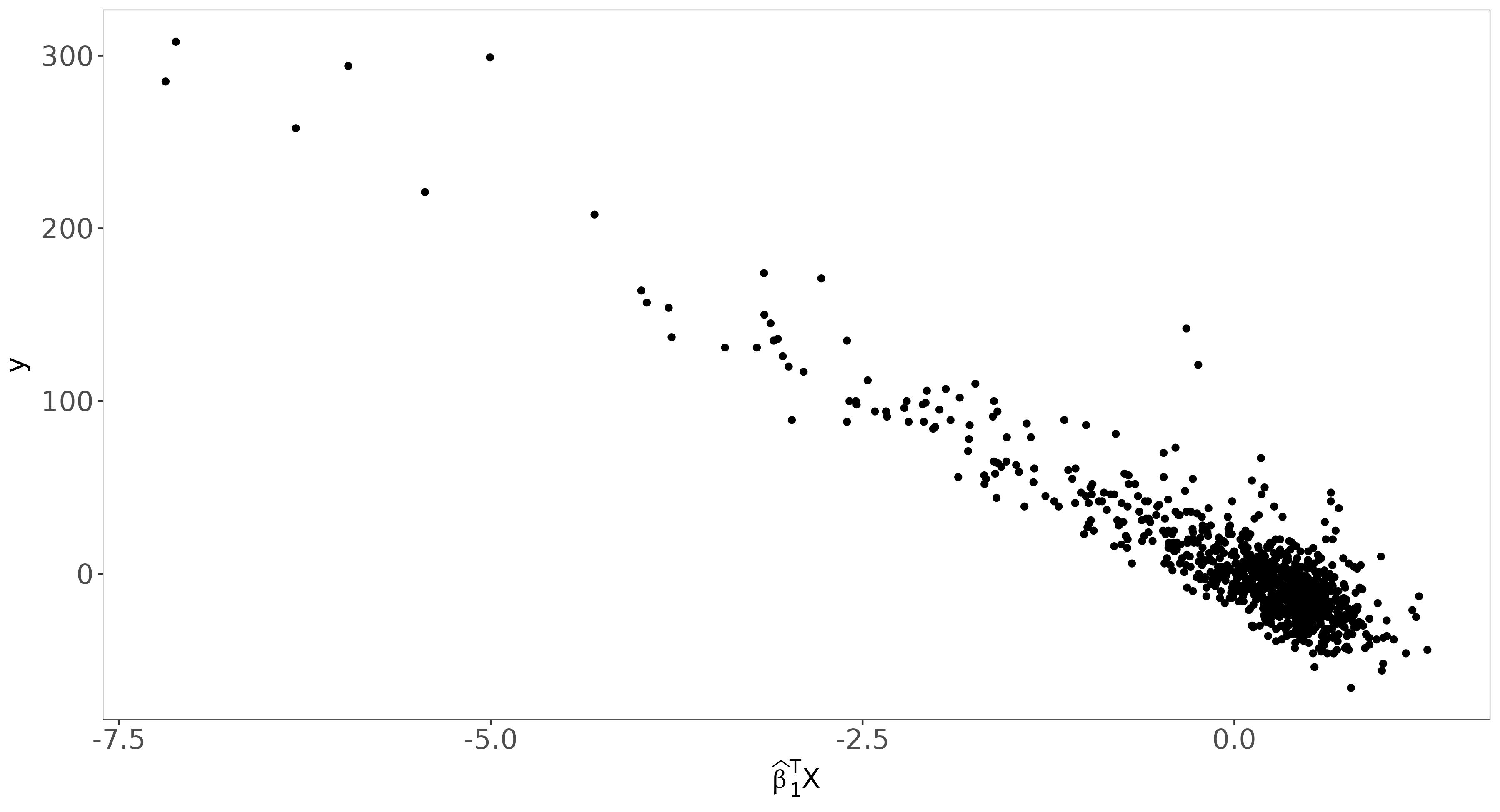}}
	\subcaptionbox{FSIR-IID}{\includegraphics[width=0.32\textwidth]{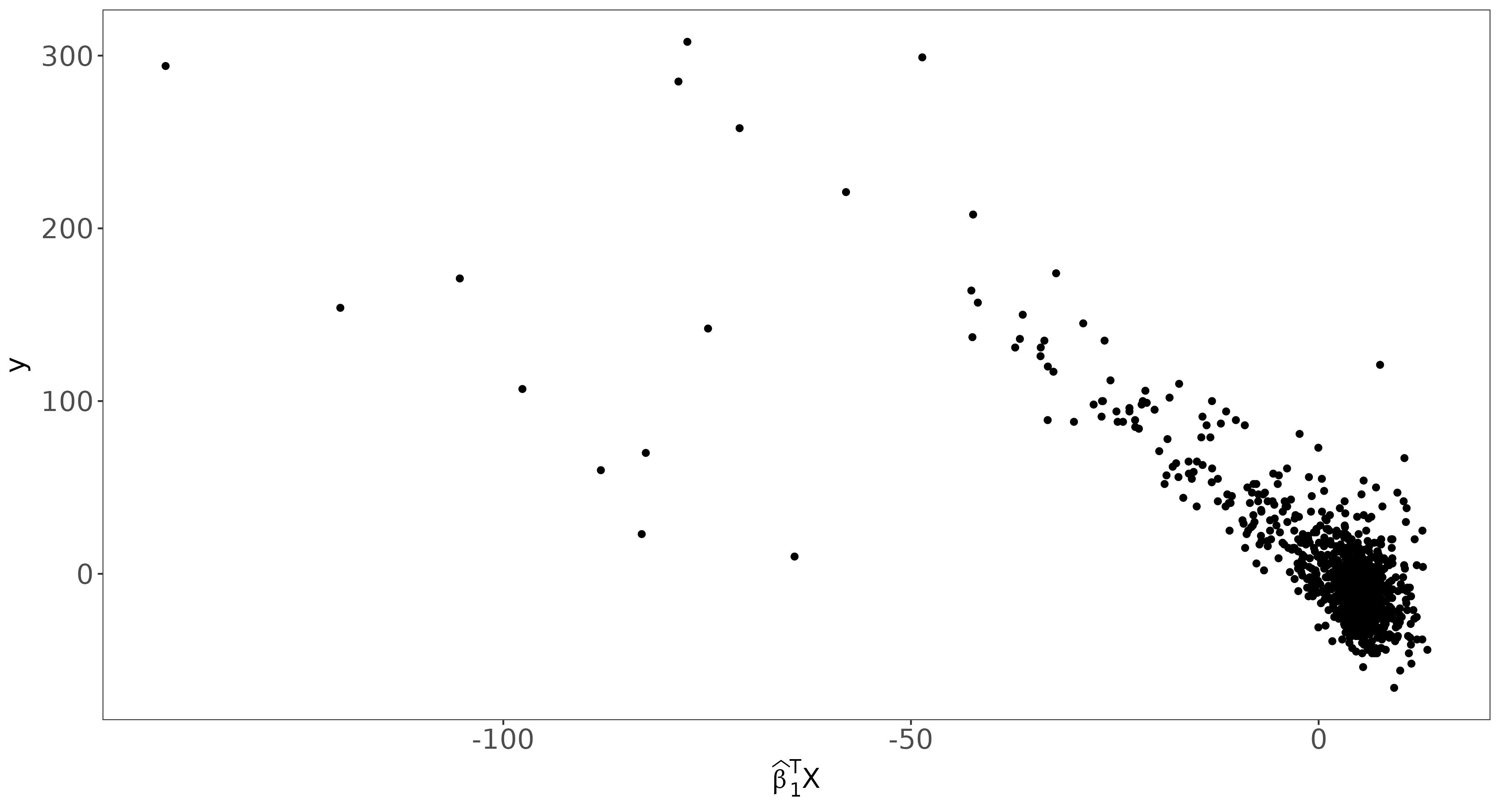}}
	\subcaptionbox{FSIR-VGM}{\includegraphics[width=0.32\textwidth]{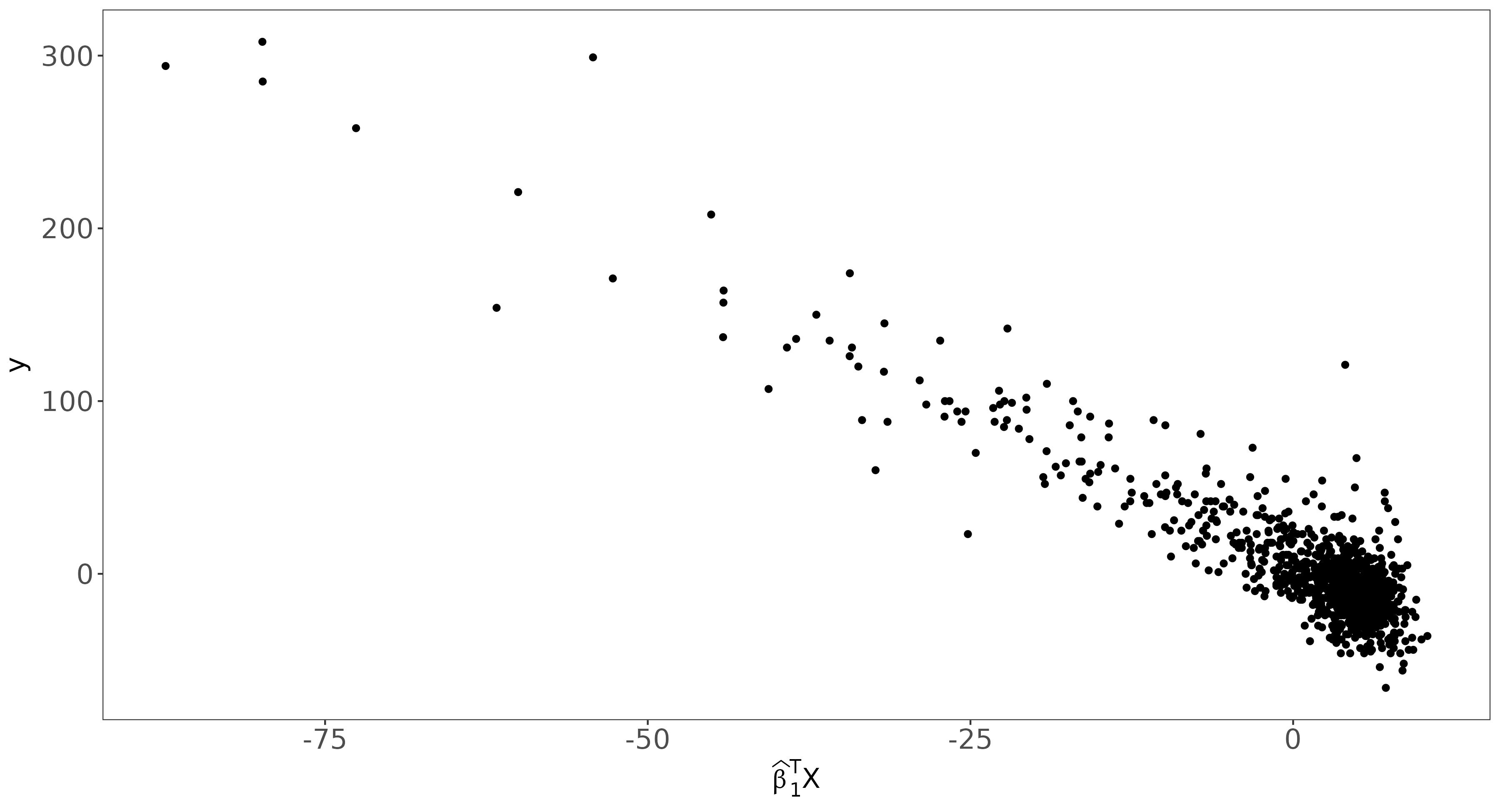}}
	\caption{Arrival Delay plotted against projected features from FSIR, FSIR-IID and FSIR-VGM on a demo subset.}
    \label{fig:real-airline}
\end{figure}

We apply our method to an airline on-time performance data in the R package ``nycflights13''. 
This dataset comprises information on $336,776$ flights that departed from various airports in NYC (e.g., EWR, JFK, and LGA) to different destinations in year 2013. 
Our object is to analyze the Arrival Delay based on seven variables: Month, Day, Departure Delay, Arrival time, Scheduled arrival time, Air time, and Distance. 
Notice the response Arrival Delay is continuous, we take $H=6$ slices. 
The whole dataset is divided into disjoint clients according to $11$ selected airlines, each containing $n=5000$ samples.  
We set the privacy level set to $(\epsilon,\delta) = (1,1/n^{1.1})$, ensuring a satisfactory level of privacy protection for the clients' data. 
Given the estimated structure dimension $\hat{d}=1$, Figure \ref{fig:real-airline} plots Arrival Delay (denoted by $y$) against the projected feature $\widehat{\beta}_1^{\T}X$, based on a randomly selected subset of $1000$ samples. 
Clearly, the VGM mechanism successfully preserves the pattern observed in FSIR.

\section{Discussion}\label{sec:final}
We provide a federated framework for performing sufficient dimension reduction on decentralized data, ensuring both data ownership and privacy.
As the first investigation on privacy protection of the SDR subspace estimation, we clarify our idea in a relative simple setting, which implicitly assumes that data on different clients are homogeneous. 
In practice, however, data might not be identically distributed across clients and we leave this heterogeneous scenario for future study. 
Considering the SIR estimator still suffers from some limitations, we claim that strategies and technical tools provided in the work can help many other SDR estimators develop their own federated extensions, with a differentially private guarantee.

\section*{Supplementary material} \label{SM}


\bibliographystyle{plainnat}
\bibliography{paper-ref}

\end{document}